%% file: WangW.tex
\documentclass{aa}
\usepackage{epsfig}
\newcommand{\double} {\baselineskip 1.5 \baselineskip}

\begin{document}
\title{A Reexamination of Electron Density Diagnostics for Ionized Gaseous 
Nebulae}
\author{
W. Wang \inst{1},
X.-W. Liu\inst{1},
Y. Zhang \inst{1}
and M. J. Barlow\inst{2}
}

\offprints{W. Wang,\\
\email{bwaw@bac.pku.edu.cn}}

\institute{Department of Astronomy, Peking University, Beijing 100871,
          P. R. China
\and Department of Physics and Astronomy, University College London, 
       Gower Street, London WC1E 6BT, U.K.
}

\date{Received??/Accepted??}
\abstract{
\double
We present a comparison of electron densities derived from optical
forbidden line diagnostic ratios for a sample of over a hundred nebulae.  We
consider four density indicators, the [\ion{O}{ii}]
$\lambda3729/\lambda3726$, [\ion{S}{ii}] $\lambda6716/\lambda6731$,
[\ion{Cl}{iii}] $\lambda5517/\lambda5537$ and [\ion{Ar}{iv}]
$\lambda4711/\lambda4740$ doublet ratios. Except for a few H ~{\sc ii} regions
for which data from the literature were used, diagnostic line ratios were
derived from our own high quality spectra.  For the [\ion{O}{ii}]
$\lambda3729/\lambda3726$ doublet ratio, we find that our default atomic data
set, consisting of transition probabilities from Zeippen (\cite{zeippen1982})
and collision strengths from Pradhan (\cite{pradhan}), fit the observations
well, although at high electron densities, the [\ion{O}{ii}]doublet ratio yields densities
systematically lower than those given by the [\ion{S}{ii}]
$\lambda6716/\lambda6731$ doublet ratio, suggesting that the ratio of
transition probabilities of the [\ion{O}{ii}] doublet,
$A(\lambda3729)/A(\lambda3726)$, given by Zeippen (\cite{zeippen1982}) may need
to be revised upwards by approximately 6 per cent.  Our analysis also shows
that the more recent calculations of [\ion{O}{ii}] transition probabilities by
Zeippen (\cite{zeippen1987a}) and collision strengths by McLaughlin \&
Bell (\cite{mclaughlin}) are inconsistent with the observations at the high and low
density limits, respectively, and can therefore be ruled out.  We confirm the
earlier result of Copetti \& Writzl (\cite{copetti2002}) that the [\ion{O}{ii}]
transition probabilities calculated by Wiese et al. (\cite{wiese}) yield
electron densities systematically lower than those deduced from the
[\ion{S}{ii}] $\lambda6716/\lambda6731$ doublet ratio and that the discrepancy
is most likely caused by errors in the transition probabilities calculated by
Wiese et al. (\cite{wiese}).  Using our default atomic data set for
[\ion{O}{ii}], we find that $N_{\rm e}([\ion{O}{ii}]) \la N_{\rm e}([\ion{S}{ii}])
\approx N_{\rm e}([\ion{Cl}{iii}])< N_{\rm e}([\ion{Ar}{iv}])$. 

\keywords{Atomic data - ISM: lines and bands - planetary nebulae: general}
}
\authorrunning{W. Wang et al.}
\titlerunning{Electron density diagnostics for ionized gaseous nebulae}
\maketitle

\section{Introduction}

Electron density, $N_{\rm e}$, is one of the key physical parameters
characterizing an ionized gaseous nebula. Accurate measurement of $N_{\rm e}$
 is a prerequisite to the determination of nebular chemical abundances and
calculation of the mass of ionized gas. Information on $N_{\rm e}$ can also be
used to estimate the distance to a nebula. The electron density in an ionized gaseous
nebula can be measured by observing the effects of collisional de-excitation on
nebular (forbidden) emission lines. This is usually achieved by comparing
the observed intensities of lines emitted from two different energy levels of
nearly equal excitation energy from the same ion, such that their intensity ratio is
insensitive to electron temperature ($T_{\rm e}$). If the two levels have very different
radiative transition probabilities, then the relative populations of the two
levels will vary with electron density, as will the intensity ratio of transitions
emitted from them.

In the optical wavelength region, the most commonly used density-diagnostic
ratios are [\ion{O}{ii}] $\lambda3729/\lambda3726$ and [\ion{S}{ii}]
$\lambda6716/\lambda6731$.  The two doublets are intrinsically strong and easy
to observe from an ionized gaseous nebula. O$^0$ and S$^0$ have similar
ionization potentials, thus one expects [\ion{O}{ii}] and [\ion{S}{ii}] lines
to arise from similar ionization regions.  The [\ion{O}{ii}] and [\ion{S}{ii}]
doublet ratios should therefore yield comparable values of electron density for
a given nebula. Other density-diagnostic ratios observable in the optical
include [\ion{Cl}{iii}] $\lambda5517/\lambda5537$ and [\ion{Ar}{iv}]
$\lambda4711/\lambda4740$. These lines are in general weaker than the
[\ion{O}{ii}] and [\ion{S}{ii}] lines and are most useful for high excitation
nebulae, in particular for planetary nebulae (PNe).  [\ion{Cl}{iii}] and
[\ion{Ar}{iv}] trace regions of higher ionization degree, thus depending on the
density structure of a nebula, densities from them need not be the same as
those derived from the low excitation [\ion{O}{ii}] and [\ion{S}{ii}] lines.
Owing to differences in their critical densities, diagnostic lines of
different ions and emitted from different pairs of energy levels are useful for
different density regimes -- [\ion{O}{ii}] and [\ion{S}{ii}] lines are
sensitive to density variations for $N_{\rm e} \sim 10^3$\,cm$^{-3}$,
[\ion{Cl}{iii}] for $N_{\rm e} \sim  10^4$\,cm$^{-3}$ and [\ion{Ar}{iv}] for
$N_{\rm e} \sim  10^4 - 10^5$\,cm$^{-3}$. Lines with relatively low critical
densities, such as the [\ion{O}{ii}] and [\ion{S}{ii}] doublets, are suppressed
in high density regions by collisional de-excitation. Thus in a highly
stratified nebula containing dense ionized clumps, [\ion{O}{ii}] and
[\ion{S}{ii}] emission will be biased towards regions of lower densities,
electron densities derived from the [\ion{O}{ii}] and [\ion{S}{ii}] doublet
ratios should therefore be lower than those yielded by the [\ion{Cl}{iii}] and
[\ion{Ar}{iv}] doublet ratios. The above simple reasoning is in consistence with 
the conclusion of the numerical analysis demonstrated by Rubin (\cite{rubin1989})
about how and to what extend does the $N_{\rm e}$ variations within a nebula affect the
 results of empirical $N_{\rm e}$-diagnostics, which show a progression in inferred 
values of $N_{\rm e}$ with the same sequence of $N_{\rm c}$ changing from the lowest
to the highest.

The low transition probabilities of nebular forbidden lines and the low impact
energies involved in a photoionized gaseous nebula imply that the atomic data
needed for the calculation of level populations and line intensity
ratios as a function of $N_{\rm e}$ (and $T_{\rm e}$), including Einstein
spontaneous transition probabilities and collision strengths, have to rely on
sophisticated multi-electron quantum mechanics calculations.  Validating the
accuracy of those theoretical calculations through comparison with observations
is thus an important aspect of nebular studies. The comparison of electron
densities derived from various diagnostics of different characteristics, such
as ionization potential, critical density and excitation energy, can provide
valuable information about nebular structure. 

In an earlier study of electron densities in PNe, Stanghellini \& Kaler
(\cite{SK1989}) compared electron densities derived from the above four optical
diagnostic ratios for a large sample using measurements published in the
literature. They found that on average the [\ion{O}{ii}] doublet ratio yielded
densities 16 per cent lower than values determined from the [\ion{S}{ii}]
doublet ratio, and about 35 per cent lower than deduced from the
[\ion{Cl}{iii}] doublet ratio. They also found that densities deduced from the
[\ion{Cl}{iii}] and [\ion{Ar}{iv}] doublet ratios were well correlated,
although the latter tended to yield slightly lower values.  They concluded that
the atomic data for these diagnostic ratios were in good shape. Using the
double-beam spectrograph mounted on the ANU 2.3m telescope, Kingsburgh \&
English (\cite{kingsburgh1992}) measured the [O~{\sc ii}] and [S~{\sc ii}]
ratios simultaneously for a large sample of Galactic PNe and found that
electron densities derived from the two ratios were in excellent agreement. On
the other hand, a very different result was obtained by Meatheringham \& Dopita
(\cite{MD1991}) who found that for a sample of 44 PNe in the Magellanic Clouds,
[\ion{S}{ii}] densities were systematically lower than those determined from
the [\ion{O}{ii}] doublet ratio. Using high resolution spectroscopic data
obtained with an echelle spectrograph, mostly by the group led by Aller (Aller
\& Hyung 1995; Aller et al.  1996; Feibelman et al. 1994, 1996; Hyung 1994;
Hyung \& Aller 1995a,~b, 1996, 1997a,~b, 1998; Hyung et al. 1993, 1994a,~b,~c
,1995, 1997, 1999a,~b, 2000, 2001; Keyes et al.  1990; Keenan et al. 1993, 1996,
1997) they determined simultaneously electron densities and temperatures
using a number of plasma diagnostic lines in the optical and ultraviolet and
found that overall the data yielded compatible and consistent results. More
recently, Copetti \& Writzl (\cite{copetti2002}), using data published in the
literature, compared the observed line ratios for a number of
density-diagnostics, instead of densities deduced from them, thus avoiding
discarding measurements that yield ratios close to or beyond the the low- or
high-$N_{\rm e}$ limits. They concluded that in general $N_e([\ion{N}{i}]) \le
N_e([\ion{O}{ii}]) < N_e([\ion{S}{ii}]$, $[\ion{Cl}{iii}], [\ion{Ar}{iv}])$ as
well as $N_e([\ion{S}{ii}]) \simeq N_e([\ion{Cl}{iii}]) \simeq
N_e([\ion{Ar}{iv}])$ and interpreted the results in terms of nebular
inhomogeneities and possible errors in atomic parameters, in particular those
of $[\ion{O}{ii}]$.

At low plasma densities, each collisional excitation by electron impact
leads to the emission of a photon. Therefore at low densities, the doublet flux ratio of
an $N_{\rm e}$-diagnostic, such as [\ion{O}{ii}] $\lambda3729/\lambda3726$
, is given by the ratio of collision strengths from the ground
$^4$S$_{3/2}$ level to the corresponding upper levels $^2$D$_{5/2}$ and
$^2$D$_{3/2}$ of the $\lambda3729$ and $\lambda3726$ lines, respectively.  It
has been commonly assumed that {\it LS}-coupling is a valid approximation for
low excitation spectral terms of ground electron configurations of light
element ions such as [\ion{O}{ii}]. Under this assumption, the ratio of
collision strengths from $^4$S$_{3/2}$ to $^2$D$_{5/2}$ and $^2$D$_{3/2}$ is
simply given by the ratio of the statistical weights of the upper two levels,
i.e. 1.50 in this case. Employing the largest set of base functions hitherto,
McLaughlin \& Bell (\cite{mclaughlin}) recently recalculated collision strengths 
for [\ion{O}{ii}] using the R-matrix method at the Breit-Pauli approximation 
level, and found that their results differ significantly from previous work 
such as those of Pradhan (\cite{pradhan}). Most surprisingly, they find that 
the ratio of collision strengths from the $^4$S$_{3/2}$ level to the $^2$D$_{5/2}$
and $^2$D$_{3/2}$ levels is not 1.50, but is about 1.93. McLaughlin \& Bell
attributed the result to departure from {\it LS}-coupling occurring in their
calculations due to relativistic effects.  This result, if confirmed, would
have a profound significant effect on nebular studies, especially $N_{\rm e}$
determinations for low surface brightness H~{\sc ii} regions which typically
have very low electron densities.  The extensive literature survey by Copetti
\& Writzl (\cite{copetti2002}) failed to find evidence in support of the result
of McLaughlin \& Bell (\cite{mclaughlin}).
\input{Table01.tex}
\input{Table02.tex}

In this paper, we present measurements of intensity ratios of the four optical
$N_{\rm e}$-diagnostics for a large sample of $> 100$ PNe. All measurements
were derived from our own high quality spectra taken with a long-slit
spectrograph equipped with a CCD detector. In Section 2 we describe the new
observations and data reduction procedures and present the results in Section
3. The new measurements are analyzed and discussed in Section 4. In order to
better constrain the low-$N_{\rm e}$ limit of the [\ion{O}{ii}] doublet ratio, we
have included in our analysis the Galactic and Magellanic Cloud H~{\sc ii}
regions studied by Tsamis et al. (\cite{tsamis}) and the extragalactic giant
H~{\sc ii} regions studied by Esteban et al. (\cite{esteban}). Diffuse
galactic emission from low-$N_{\rm e}$ ionized gas offers the best medium to
constrain observationally the low-$N_{\rm e}$ limit of the [\ion{O}{ii}] and
[\ion{S}{ii}] doublet ratios. Measurements of such diffuse emission available
in the literature are therefore discussed as well.

\section{Observation and data reduction}

Most of the observations were obtained with the ESO 1.52\,m telescope using the
Boller \& Chivens (B\&C) long-slit spectrograph in three observing runs in
1995, 1996 and 2001.  In 1995, the spectrograph was equipped with a Ford
2048$\times$2048 15$\mu$m$\times$15$\mu$m CCD, which was superseded in 1996 by
a UV-enhanced Loral $2048\times2048$ $15\,\mu{\rm m}\times 15\,\mu{\rm m}$ chip
and in 2001 by a Loral $2688\times 2688$ $15\,\mu{\rm m}\times 15\,\mu{\rm m}$
chip.

A number of northern hemisphere PNe were later observed with the 2.5\,m Isaac
Newton Telescope (INT) using the Intermediate Dispersion Spectrograph (IDS) and
the 235\,mm camera equipped with an EEV CCD. The spectral wavelength coverage, slit
width and FWHM resolution for each observing run are listed in
Table~\ref{jan}.

Typical exposure times were about 5\,min for low resolution spectra which had a
FWHM of about 4.5\,{\AA} and about 20 to 30\,min for spectra of higher
resolution. Most nebulae were observed with a fixed position long-slit,
sampling the brightest parts of the nebulae, normally passing through the nebular
centre. For a few nebulae, such as NGC\,6153, scanned spectra were also
obtained by uniformly driving a long slit across the nebular surface, thus
yielding average spectra for the entire nebula.

All spectra were reduced using the {\sc long92} package in {\sc
midas}\footnote{{\sc midas} is developed and distributed by the European
Southern Observatory.} following the standard procedure. Spectra were
bias-subtracted, flat-fielded and cosmic-rays removed, and then wavelength
calibrated using exposures of a calibration lamp. 

\input{Table03.tex}

Given that in the current work, we are interested only in intensity ratios of
two lines close in wavelength, the effects of any uncertainties in flux
calibration and corrections for interstellar dust extinction on our results are
minimal and can be safely neglected. Uncertainties in the observed doublet
ratios are therefore dominated by the photon counting noise for the lines
involved. The [\ion{S}{ii}] $\lambda\lambda$6716, 6731 and [\ion{O}{ii}]
$\lambda\lambda$3726, 3729 lines are generally quite strong and easy to
measure, except for some optically thin, high excitation PNe. The
[\ion{Cl}{iii}] $\lambda\lambda$5517, 5537 lines are intrinsically much weaker,
given the lower chlorine abundance. The [\ion{Ar}{iv}]
$\lambda4711/\lambda4740$ lines can also be quite weak, especially in low
excitation PNe. In our spectra, the [\ion{Ar}{iv}] 4711.37\AA\ line was
partially blended with the He~{\sc i} 4713.17\AA\ line even for those spectra
taken with a spectral resolution of 1.5\,{\AA} FWHM. In PNe of very high
excitation class, the [Ne~{\sc iv}] $\lambda\lambda$4714.25,4715.61 lines may
also contribute to the observed flux of the 4712\AA\ feature. On the other
hand, in such cases, the intensities of the $\lambda\lambda$4714.25,4715.61
lines can be well constrained using the other two [Ne~{\sc iv}] lines at
4724.15 and 4725.62\AA\ arising from the same upper level $^2$P$_{1/2,3/2}$.
Using this, together with the fact that the laboratory wavelengths of all these
lines are well known, we found that in most cases, the intensity of the
[Ar~{\sc iv}] $\lambda$4711 line can be retrieved using Gaussian line profile
fitting. The formal errors as given by the line fitting program were then
propagated into the line ratios and electron densities deduced from them
following standard procedures. The uncertainties could be underestimated or
overestimated, depending on the relative intensities of the blending
[\ion{Ar}{iv}] $\lambda$4711 and He~{\sc i} $\lambda$4713 lines.  

As a further check of the reliability and accuracy of our line profile fitting
technique, we have compared for a number of sample PNe (e.g. Cn\,2-1,
He\,2-118) line fluxes derived from the current data set with those measured on
spectra newly obtained under a higher resolution ($\sim 0.9$\AA\ FWHM).  The
agreement between the two sets of measurement were found to be quite
satisfactory, with typical differences of less than 3\% when
I($\lambda4713$)/I($\lambda4711$) ratio is about 2, and about 0.2\,\% when the
ratio is about 0.5. The errors introduced by line profile fitting procedures
are therefore well within the photon counting noise limits.

For two sample PNe (M\,2-24 and IC\,4406), only low resolution spectra (FWHM =
4.5\AA) were available.  For these two nebulae contributions from the He~{\sc
i} $\lambda$4713 line and the [Ne~{\sc iv}] $\lambda\lambda$4714,4716 lines (as
in the case of the high excitation PN IC\,4406) to the $\lambda$4711 feature
were estimated from the observed intensities of the He~{\sc i} $\lambda$4471
and [Ne~{\sc iv}] $\lambda\lambda$4724,4726, respectively, and subtracted.
Further observations of better spectral resolution should be useful.

The [\ion{Cl}{iii}] $\lambda\lambda$5517,5537 lines were only marginally
detected in Sn\,1 and NGC\,6072, resulting fairly large uncertainties in the
line fluxes and their ratio.

\section{Results}

%fig1
\begin{figure}
\centering 
\epsfig{file=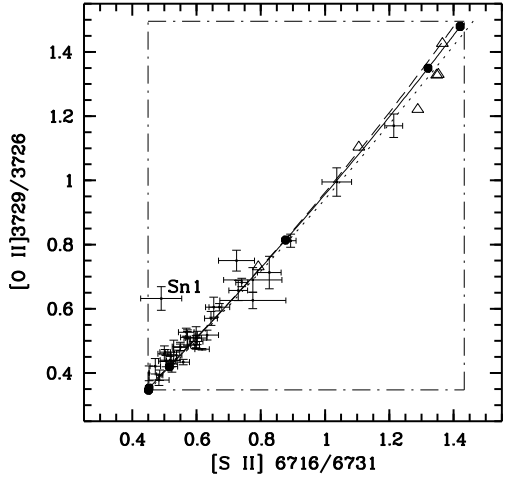, height=6.5cm,
bbllx=55, bblly=70,bburx=300, bbury=302, clip=, angle=0}
\caption{
The observed [\ion{O}{ii}] $\lambda3729/\lambda3726$ ratio is plotted against 
the [\ion{S}{ii}] $\lambda6716/\lambda6731$ ratio for 37 nebulae. Triangles are
H~{\sc ii} regions from Tsamis et al. (\cite{tsamis}) and Esteban et al.
(\cite{esteban}).  The dotted, solid and dashed lines delineate the theoretical
variations of the [\ion{O}{ii}] ratio as a function of the [\ion{S}{ii}] ratio,
assuming that both ions arise from identical ionization regions of uniform
density, for an $T_{\rm e}$ of 5\,000~K, 10\,000~K and 15\,000~K,
respectively. Along the lines, $N_{\rm e}$ increases from top right to
bottom left, and the six filled circles denote densities of $10^0$, $10^2$, 
$10^3$, $10^4$, $10^5$, $10^6$\,cm$^{-3}$, respectively. The dotted-dashed 
rectangle indicates the region of allowed values of the two diagnostic ratios 
between the low- and high-density limits. 
}
\label{f1} 
\end{figure}

%fig2
\begin{figure}
\centering
\epsfig{file=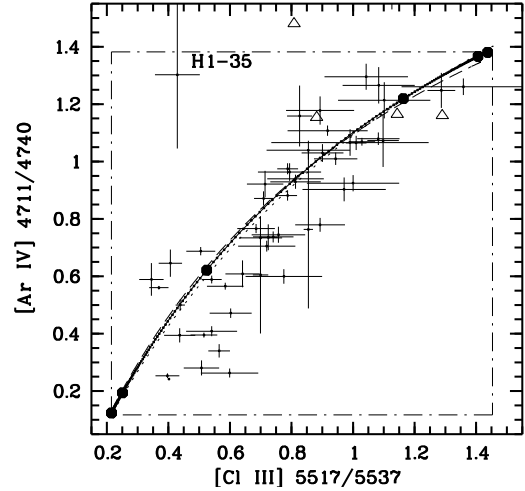, height=6.5cm,
bbllx=60, bblly=75, bburx=300, bbury=300, clip=, angle=0}
\caption{Same as Fig.\,1 but for [\ion{Ar}{iv}] $\lambda4711/\lambda4740$
against [\ion{Cl}{iii}] $\lambda5517/\lambda5537$. Observed values for a total 
of 59 nebulae are plotted. }
\label{f2}
\end{figure}

%fig3
\begin{figure}
\centering
\epsfig{file=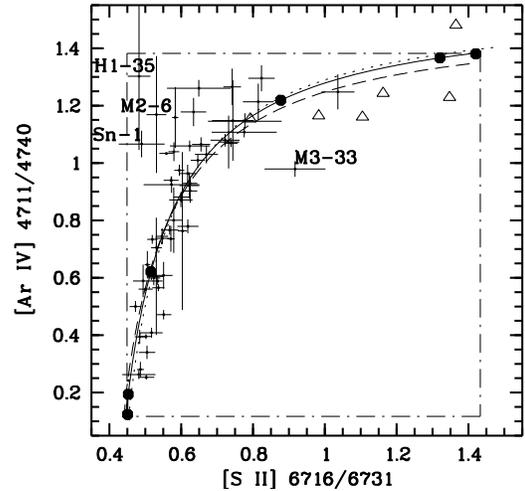, height=6.5cm,
bbllx=60, bblly=75, bburx=300, bbury=300, clip=, angle=0}
\caption{Same as Fig.\,1 but for [\ion{Ar}{iv}] $\lambda4711/\lambda4740$ 
against [\ion{S}{ii}] $\lambda6716/\lambda6731$. 62 nebulae are plotted. }
\label{f3}
\end{figure}
 
%fig4
\begin{figure}
\centering
\epsfig{file=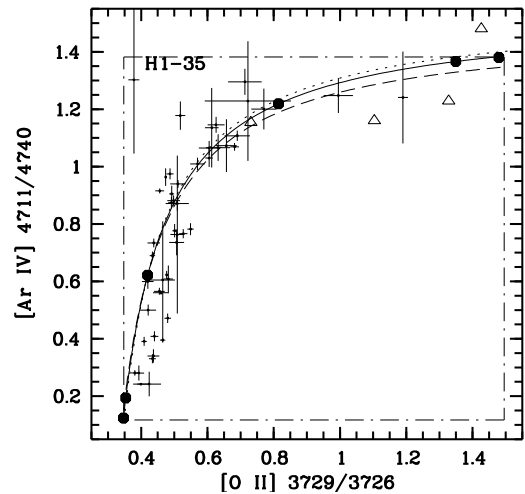, height=6.5cm,
bbllx=60, bblly=75, bburx=300, bbury=300, clip=, angle=0}
\caption{Same as Fig.\,1 but for [\ion{Ar}{iv}] $\lambda4711/\lambda4740$
plotted against [\ion{O}{ii}] $\lambda3729/\lambda3726$ for 36 nebulae. }
\label{f4}
\end{figure}

%fig5
 \begin{figure}
 \centering
 \epsfig{file=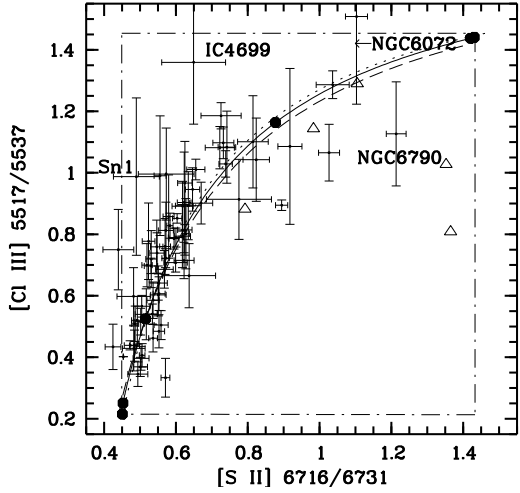, height=6.5cm,
bbllx=60, bblly=75, bburx=300, bbury=300, clip=, angle=0}
\caption{Same as Fig.\,1 but for [\ion{Cl}{iii}] $\lambda5517/\lambda5537$
against [\ion{S}{ii}] $\lambda6716/\lambda6731$ for 74 nebulae. Note that 
in this digram the two filled circles at the top-right representing 
$N_{\rm e} = 10^0$ and $10^2$ cm$^{-3}$ fall almost on top of each other.}
\label{f5}
\end{figure}

%fig6
 \begin{figure}
 \centering
 \epsfig{file=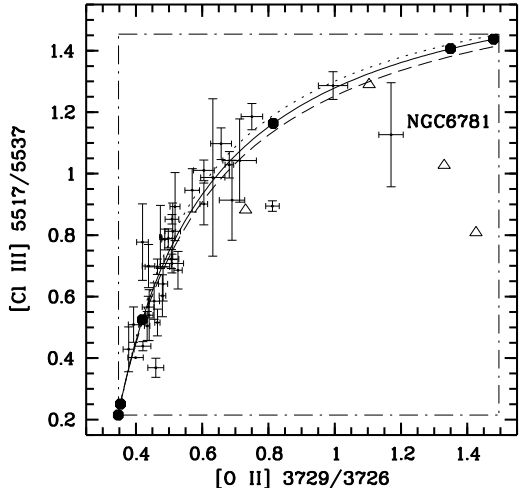, height=6.5cm,
bbllx=60, bblly=75, bburx=300, bbury=300, clip=, angle=0}
\caption{Same as Fig.\,1 but for [\ion{Cl}{iii}] $\lambda5517/\lambda5537$
against [\ion{O}{ii}] $\lambda3729/\lambda3726$ for
36 nebulae. }
\label{f6}
\end{figure} 

%fig7
 \begin{figure}
 \centering
 \epsfig{file=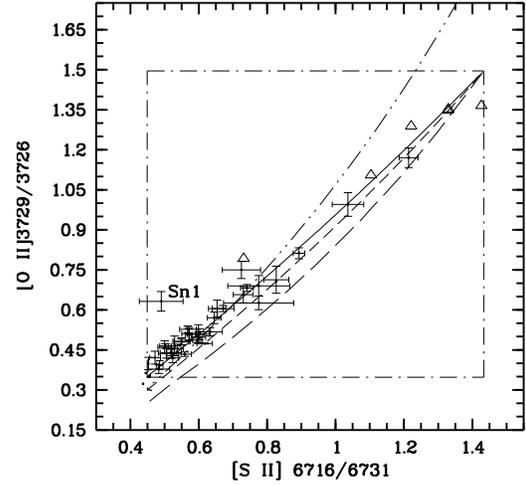, height=6.5cm,
 bbllx=60, bblly=72, bburx=314, bbury=315, clip=, angle=0}
\caption{Same as Fig.\,1 except that the loci delineating variations of
the [\ion{O}{ii}] ratio as a function of the [\ion{S}{ii}] ratio for a 
homogeneous nebula were derived using combinations of atomic parameters 
tabulated in Table~\ref{fig7_ref}, where the different line types are described.
 A constant electron temperature of 10\,000~K was assumed. The six circles 
denoting different electron densities are not plotted here.
}
\label{f7}
\end{figure}
\input{Table04.tex}

%%%%fig8
\begin{figure}
\centering 
\epsfig{file=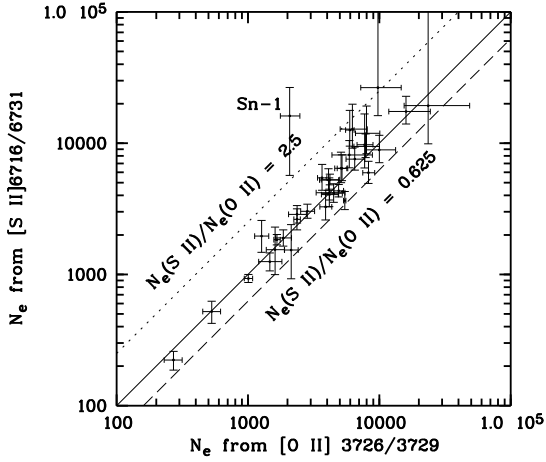, height=6.5cm, bbllx=95,
bblly=73, bburx=379, bbury=319, clip=, angle=0} 
\caption{Electron densities derived from the [\ion{S}{ii}] 
$\lambda6716/\lambda6731$ ratio plotted against value derived from the
[\ion{O}{ii}] $\lambda3729/\lambda3726$ ratio. The dotted and dashed
diagonal lines represent [\ion{S}{ii}] to [\ion{O}{ii}] $N_{\rm e}$ ratios 
of 2.5 and 0.625, respectively.} 
\label{f8}
\end{figure}

%%%%fig9
\begin{figure}
\centering
\epsfig{file=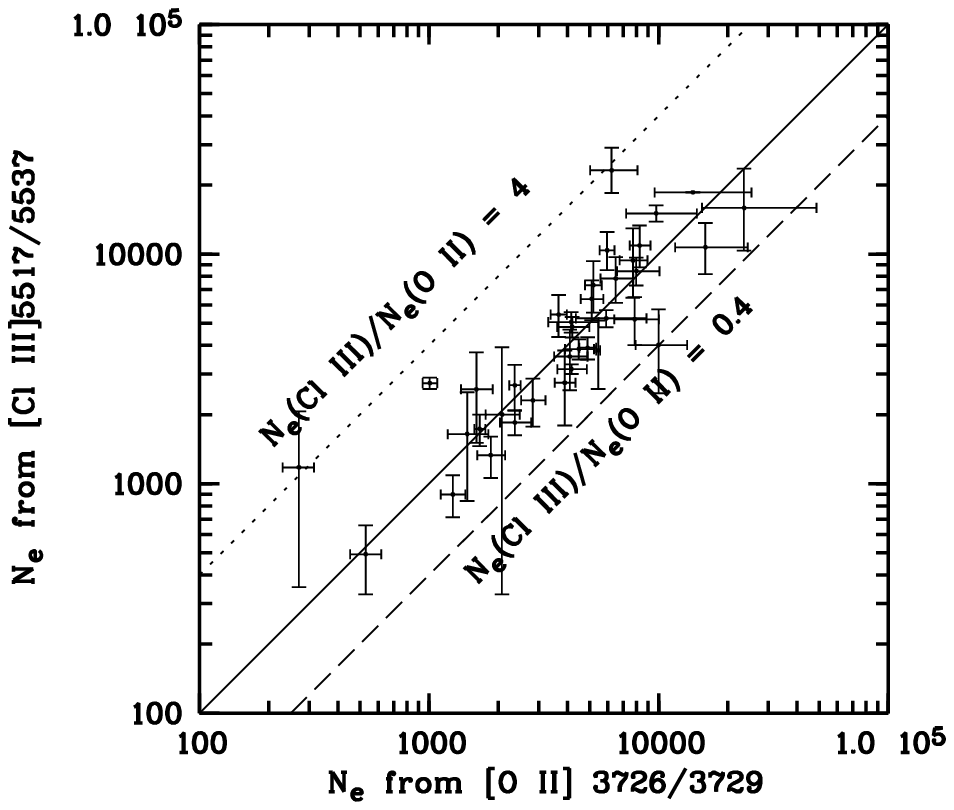, height=6.5cm,
bbllx=95, bblly=73, bburx=379, bbury=309, clip=, angle=0}
\caption{Same as Fig.\,8 but for densities derived from the
[\ion{Cl}{iii}] and [\ion{O}{ii}] doublet ratios, respectively. }
\label{f9}
\end{figure}

%%%%fig10
\begin{figure}
\centering
\epsfig{file=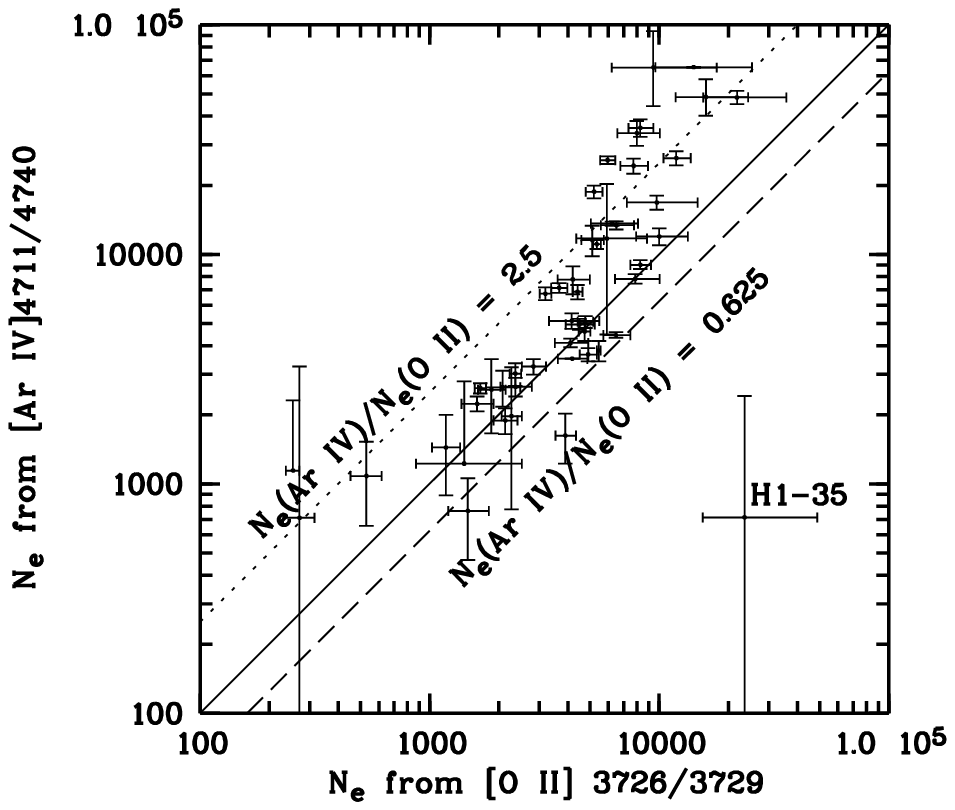, height=6.5cm,
bbllx=70, bblly=60, bburx=347, bbury=296, clip=, angle=0}
\caption{Same as Fig.\,8 but for densities derived from the
[\ion{Ar}{iv}] and [\ion{O}{ii}] doublet ratios, respectively. }
\label{f10}
\end{figure}

%%%%fig11
\begin{figure}
\centering
\epsfig{file=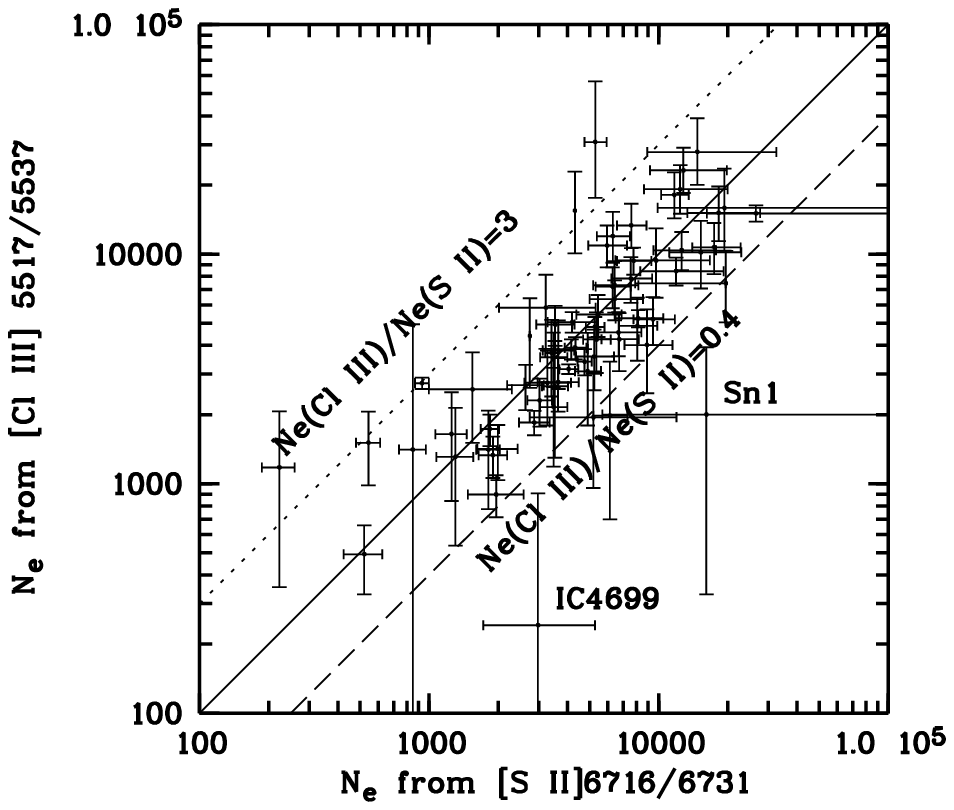, height=6.5cm,
bbllx=98, bblly=72, bburx=380, bbury=308, clip=, angle=0}
\caption{Same as Fig.\,8 but for densities derived from the
[\ion{Cl}{iii}] and [\ion{S}{ii}] doublet ratios, respectively. }
\label{f11}
\end{figure}

%%%%fig12
\begin{figure}
\centering
\epsfig{file=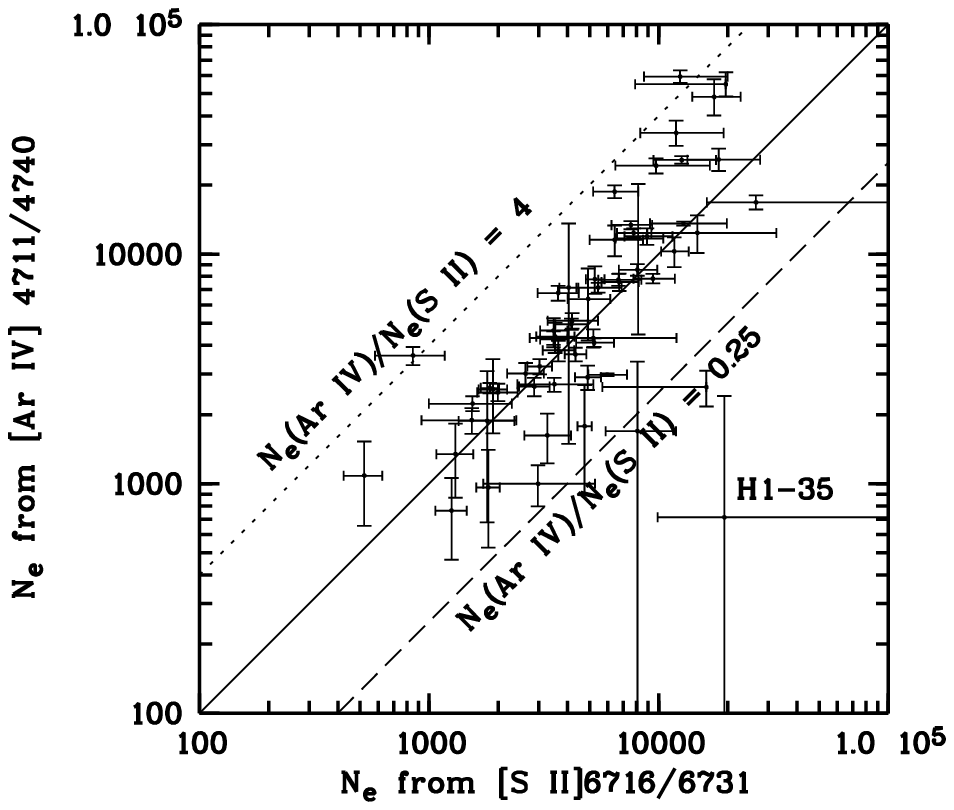, height=6.5cm,
bbllx=98, bblly=72, bburx=380, bbury=308, clip=, angle=0}
\caption{Same as Fig.\,8 but for densities derived from the
[\ion{Ar}{iv}] and [\ion{S}{ii}] doublet ratios, respectively. }
\label{f12}
\end{figure}

%%%%fig13
\begin{figure}
\centering
\epsfig{file=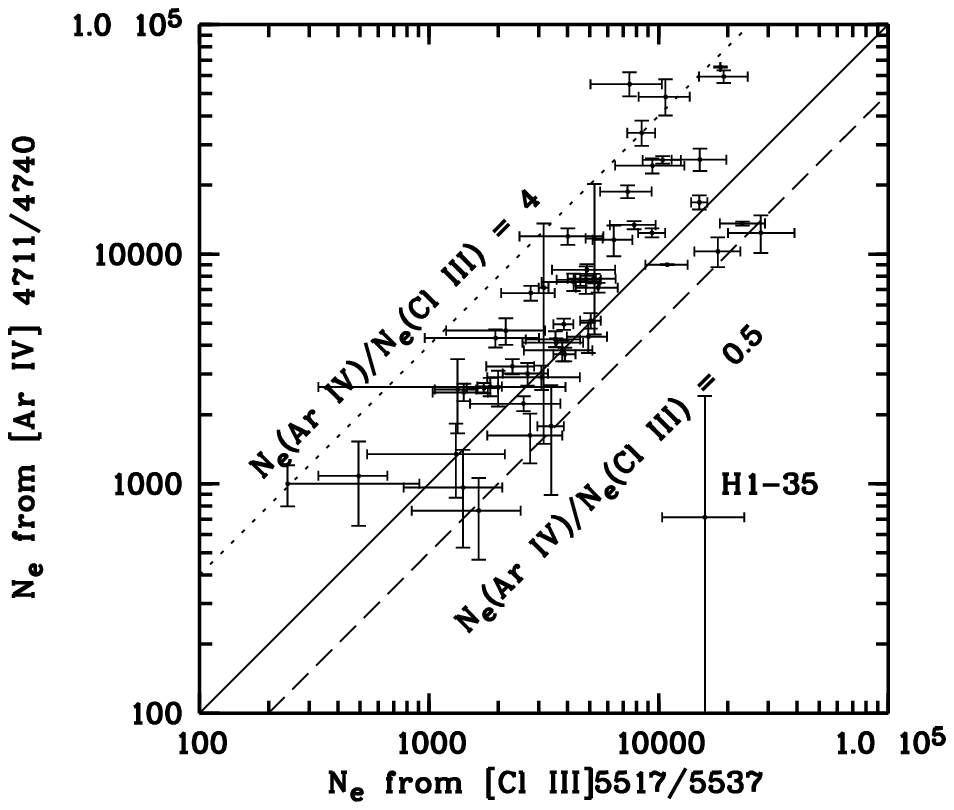, height=6.5cm,
bbllx=95, bblly=73, bburx=375, bbury=309, clip=, angle=0}
\caption{Same as Fig.\,8 but for densities derived from the
[\ion{Ar}{iv}] and [\ion{Cl}{iii}] doublet ratios, respectively.}
\label{f13}
\end{figure}

Intensity ratios for the four optical $N_{\rm e}$-diagnostic doublets, derived from
our new observations, are presented in Table~\ref{source}, along with their
uncertainties. Electron densities deduced from the ratios are also given in the
Table, all derived using {\sc equib}, a Fortran code to solve level
populations for a multi-level atom, assuming a nominal electron temperature of
10\,000~K. A five level atomic model was assumed for the four ions of interest
here. References for our default atomic data set, including collision strengths
and transition probabilities, are listed in Table~\ref{atomic_ref}.  Also given
in the Table are ionization potentials for the ions and their predecessors.

In Figs.\,1-6 we plot against each other the observed $N_{\rm e}$-diagnostic ratios, 
[\ion{O}{ii}] $\lambda3729/\lambda3726$, [\ion{S}{ii}]$\lambda6716/\lambda6731$,
[\ion{Cl}{iii}] $\lambda5517/\lambda5537$ and [\ion{Ar}{iv}] $\lambda4711/\lambda4740$ 
against one another.  Lines over-plotted in each graph show
loci tracing variations of one diagnostic ratio as a function of another,
assuming that both ions arise from identical ionization regions of uniform
density, for the cases of $T_{\rm e}$ of 5000, 10\,000 and 15\,000~K.
To further expand our sample into the low density regime, we have included in
our analysis measurements of several Galactic and Magellanic Cloud H~{\sc ii}
regions observed by Tsamis et al. (\cite{tsamis}) and extragalactic giant H~{\sc ii}
regions observed by Esteban et al. (\cite{esteban}).  The default atomic
parameters described in Table~\ref{atomic_ref} were used.  The rectangle in
each diagram delineates the region of allowed values of the diagnostic line
ratios between their low- density(top right) and high-$N_{\rm e}$ (bottom left) limits. 

In Fig.\,7, the [\ion{O}{ii}] $\lambda3729/\lambda3726$ ratio is plotted
against the [\ion{S}{ii}] $\lambda6716/\lambda6731$ ratio and compared to
theoretical predictions for various atomic data sets for O$^+$, including
the more recent collision strengths of McLaughlin \& Bell (\cite{mclaughlin})
who found that the ratio of collision strengths from the ground $^4$S$_{3/2}$
level to the $^2$D$_{5/2}$ and $^2$D$_{3/2}$ levels deviates significantly from
the {\it LS}-coupling value. Again, we assume here that the [\ion{O}{ii}]
and [\ion{S}{ii}] lines arise from identical ionization regions and that
there are no significant density inhomogeneities.

Electron densities derived from the four optical $N_{\rm e}$-diagnostics {using
the default atomic parameters listed in Table~\ref{atomic_ref}} are compared
against each other in Figs.~\ref{f8}, \ref{f9}, \ref{f10}, \ref{f11}, \ref{f12}
and \ref{f13}. In all cases, a constant nominal $T_{\rm e}$ of 10\,000~K is
assumed in calculating the densities.  In each diagram, nebulae that deviate
significantly from the loci are labelled.

\section{Discussion}
 
At low densities [$N_{\rm e} << N_{\rm c}(\lambda; T_{\rm e})$, where $N_{\rm
c}(\lambda; T_{\rm e})$ is the critical density of the diagnostic line; c.f.
Osterbrock~\cite{osterbrock2nd}], each collisional excitation by an 
electron leads to the emission of a photon. Given the small wavelengths
interval, the energy difference can be ignored, thus the intensity ratio of an
$N_{\rm e}$-diagnostic doublet, such as [\ion{O}{ii}] $\lambda3729/\lambda3726$, 
is given by the ratio of collision strengths from
the ground $^4$S$_{3/2}$ level to the corresponding $^2$D$_{5/2}$ and
$^2$D$_{3/2}$ upper levels. Since the ground term $^4$S has only one
fine-structure level, under {\it LS}-coupling, this ratio is given by the ratio
of the statistical weights of the upper levels and equals 1.50. Similar
relations hold for the [\ion{S}{ii}], [\ion{Cl}{iii}] and [\ion{Ar}{iv}]
doublet ratios. Note that the two [\ion{O}{ii}] $^2$D fine-structure levels
are inverted, but those of [\ion{S}{ii}], [\ion{Cl}{iii}] and [\ion{Ar}{iv}]
are not. Hence for the latter three ions, the ratio of the intensity of
the shorter wavelength line to that of the longer wavelength line is 1.5, not
the reverse as in the case of [\ion{O}{ii}].

At high electron densities [$N_{\rm e} >> N_{\rm c}(\lambda; T_{\rm e})$],
collisional processes dominate over radiative decays and the upper two
$^2$D$_{3/2,5/2}$ fine-structure levels are thermalized and populated according
to their statistical weights, leading to a doublet intensity ratio that is
given by the ratio of statistical weight ($\omega$) times transition
probability (A-value). For example, in the case of [\ion{O}{ii}], at high 
densities,
\[
\frac{I(\lambda3729)}{I(\lambda3726)} \propto
\frac{\omega(^2D_{5/2})}{\omega(^2D_{3/2})} \cdot
\frac{A(^2D_{5/2}-{^4S})}{A(^2D_{3/2}-{^4S})}, 
\] 
where $\omega(^2D_{5/2})/\omega(^2D_{3/2}) = 1.5$. 

Measurements of doublet ratios near low- and high-$N_{\rm e}$ limits can
therefore be used to constrain and test the accuracy of theoretical
calculations of the relative collision strengths and spontaneous transition
probabilities of $N_{\rm e}$-diagnostic ratios. In the following, we will
discuss each of the optical doublets and compare densities derived from them.
Kolmogorov-Smirnov two-sample tests are carried out to test the null hypothesis
that densities derived from individual $N_{\rm e}$-diagnostics are consistent
with each (densities derived from two diagnostic ratios are drawn from the same
sample). 

\subsection{[\ion{O}{ii}] and [\ion{S}{ii}]}

O$^0$ and S$^0$ have comparable ionization potentials. In addition, the
[\ion{O}{ii}] and [\ion{S}{ii}] doublet lines have very similar critical
densities\footnote{For $T_{\rm e} = 10^4$~K and our default atomic parameters,
the [\ion{O}{ii}] $^2$D$_{5/2}$ and $^2$D$_{3/2}$ levels have an critical
density of 1200 and 4300\,cm$^{-3}$, respectively. The corresponding values for
the corresponding two levels of [\ion{S}{ii}] are 1400 and 3600\,cm$^{-3}$.}
and are therefore sensitive to $N_{\rm e}$ variations over a similar density
regime.  Figs.~\ref{f1} and \ref{f8} show that under the reasonable assumption
that [\ion{O}{ii}] and [\ion{S}{ii}] lines arise from similar ionization
regions, and that the nebula is more or less homogeneous, then for our default
atomic data set the measured [\ion{O}{ii}] and [\ion{S}{ii}] ratios are
consistent with each other over a wide $N_{\rm e}$ range, from less than
100\,cm$^{-3}$ to over $10^{4}$\,cm$^{-3}$. At densities higher than $\sim
10^4$\,cm$^{-3}$, there is however some evidence that the [\ion{O}{ii}] doublet
ratio indicates densities systematically lower than implied by the
[\ion{S}{ii}] ratio. Our result thus differs slightly from the earlier [O~{\sc
ii}] and [S~{\sc ii}] density survey by Kingsburgh \& English
(\cite{kingsburgh1992}). From the integrated [O~{\sc ii}] and [S~{\sc ii}]
doublet ratios measured from a homogeneous spectral data set for 64 Galactic
PNe, Kingsburgh \& English found that ${\rm log} N_{\rm e}$([S~{\sc ii}])$=
-0.31 + 1.08\times {\rm log} N_{\rm e}$([O~{\sc ii}]) and concluded that the
result was consistent with a slope of unity, indicating equality between the
two densities for their sample PNe. For [S~{\sc ii}], Kingsburgh \& English
used the collision strengths and level energies from Mendoza
(\cite{mendoza1983}). We have recalculated [S~{\sc ii}] densities from their
published line ratios using our default atomic data whereby we have adopted the
more recent calculations by Keenan et al. (\cite{keenan1996}) for [S~{\sc ii}].
The results lead to revised relation between the [O~{\sc ii}] and [S{\sc ii}]
densities, ${\rm log} N_{\rm e}$([S~{\sc ii}])$= -0.38 + 1.11\times {\rm log}
N_{\rm e}$([O~{\sc ii}]).  For comparison, our sample PNe yield ${\rm log}
N_{\rm e}$([S~{\sc ii}])$= -0.49 + 1.16\times {\rm log} N_{\rm e}$([O~{\sc
ii}]).  The slopes of the latter two relations are almost identical. We
therefore conclude that the two spectroscopic surveys for the [O~{\sc ii}] and
[S~{\sc ii}] doublet ratios are consistent with each other.

\subsubsection{Low density limit}

Copetti \& Writzl (\cite{copetti2002}) compared the observed [O~{\sc ii}]
$\lambda$3729/$\lambda$3726 doublet ratios and the [S~{\sc ii}]
$\lambda$6716/$\lambda$6731 ratios for a large sample of PNe and found that the
locus calculated using the [O~{\sc ii}] collision strengths of McLaughlin \&
Bell (\cite{mclaughlin}) disagrees with observations near the low-$N_{\rm e}$
limit. Based on this, they concluded that the results of McLaughlin \& Bell
(\cite{mclaughlin}) are probably spurious.  A similar plot is shown in
Fig.\,\ref{f7} using our own data. As noted by Copetti \& Writzl
(\cite{copetti2002}), the locus calculated using the [\ion{O}{ii}] collision
strengths of McLaughlin \& Bell deviates from observations at low densities.
[S~{\sc ii}] has a larger atomic mass number, and has a larger fine-structure
splitting than [\ion{O}{ii}], thus one would expect [\ion{S}{ii}] to be
affected more by relativistic effects.  Consequently the [\ion{S}{ii}] doublet
ratio ought to show an even larger deviation from the ratio of statistical
weights should the effects invoked by McLaughlin \& Bell (\cite{mclaughlin})
apply.  Were there to be a comparable deviation of the [S~{\sc ii}] doublet
ratio from its {\it LS}-coupling value as in the case of [O~{\sc ii}] found by
McLaughlin \& Bell (\cite{mclaughlin}), then the dash-dotted line in
Fig.~\ref{f7} can in principle be brought into agreement with observations.
Thus the apparent disagreement noticed by Copetti \& Writzl
(\cite{copetti2002}) and shown here in Fig.\,\ref{f7} is not sufficient to
reject the results of McLaughlin \& Bell (\cite{mclaughlin}).  Alternatively,
if relativistic effects are important for [S~{\sc ii}] but not for [O~{\sc
ii}], then the dotted long-dashed locus in Fig.~\ref{f7} will curve downwards
as the low-$N_{\rm e}$ limit is approached (top-right corner of the graph) -- a
prediction not supported by the observations as well.

\input{Table05.tex}

The key evidence against the results of McLaughlin \& Bell (\cite{mclaughlin})
remains the fact that none of the nebulae in our sample, including PNe and
H~{\sc ii} regions, shows a line ratio in excess of 1.5 (Fig.\,1).  The largest
$\lambda$3729/$\lambda$3726 ratio determined for a PN in our sample is
$1.23\pm0.01$ (M\,1-64, c.f. Table~1), and that for an H~{\sc ii} region is
$1.43$ (SMC\,N66, Tsamis et al. 2003). Amongst the 64 Galactic PNe observed by
Kingsburgh \& English(\cite{kingsburgh1992}), IC~5148-50 has the largest
$\lambda$3729/$\lambda$3726 ratio, $1.45\pm0.04$. Similarly, none of the
nebulae in the extensive literature survey by Copetti \& Writzl
(\cite{copetti2002}) shows a ratio approaching 1.92. 

The best medium from which to determine observationally the low-$N_{\rm e}$
limit for [O~{\sc ii}] $\lambda$3729/$\lambda$3726 is the low-$N_{\rm e}$
plasma responsible for the diffuse galactic emission.  Monk et al.
(\cite{MBC1990}) obtained deep long-slit spectra centered on and around the PNe
IC\,418 and symbiotic star RX Pup with the Anglo-Australian Telescope (AAT).
[\ion{O}{ii}] $\lambda\lambda$3729, 3726 doublet emission was detected all the
way out to an angular distance of 180 arcsec from the center of IC\,418. The
$\lambda3729$/$\lambda3726$ ratio was found to change from a minimum
$0.339\pm0.028$ to a maximum $1.78\pm0.67$, both values remain consistent with
the theoretical high- and low-$N_{\rm e}$ limits of 0.350 and 1.50,
respectively, when taking into account the observational uncertainties.
Similar results were found for RX\,Pup. There are also some other published
observations near the density limits, yet none of them show clear evidence of
inconsistency with the standard value when taking error bars into account. In
Table~\ref{oiirat} we list all the recent measurements we are aware of that
yield an [O~{\sc ii}] $\lambda$3729/$\lambda$3726 ratio in excess of 1.4.
Clearly the collision strengths of Pradhan (\cite{pradhan}) are widely
supported by observations; that is to say, {\it LS}-coupling is a good
approximation for the five lowest levels of [\ion{O}{ii}].  To our knowledge,
there are no direct observations of PNe and H~{\sc ii} regions in favor of the
collision strengths of McLaughlin \& Bell (\cite{mclaughlin}).

Similarly for [S~{\sc ii}], none of the recent high quality measurements for a
PN or an H~{\sc ii} region of the $\lambda6716$/$\lambda6731$ doublet ratio
yield an value in excess the canonical value of 1.5 under {\it LS}-coupling.
For our sample PNe, the largest ratio recorded is $1.21\pm0.03$ whereas for
those observed by Kingsburgh \& English (\cite{kingsburgh1992}) the maximum
value is $1.35\pm0.04$. In 23 of their long-slit PN spectra, Kingsburgh \&
English detected [S~{\sc ii}] $\lambda\lambda6716,6731$ emission extending all
along the slit length. Ten of the those extended emission regions show a
$\lambda6716$/$\lambda6731$ ratio in excess of 1.4 and these are listed in
Table~\ref{siirat}. Again, no values exceed the canonical value when taking
error bars into account.
 
We note that for the very low surface brightness diffuse emission regions
around PNe observed by Kingsburgh \& English (\cite{kingsburgh1992}), [S~{\sc
ii}] densities tend to be lower compared to [O~{\sc ii}] values. Given that
none of the measured [O~{\sc ii}] and [S~{\sc ii}] doublet ratios exceeds 1.5
when taking error bars into account, it seems unlikely this is caused by
effects of possible departure from {\it LS}-coupling. A plausible explanation
is that the low density Galactic diffuse gas, ionized by soft diffuse
background radiation, emits more strongly in [S~{\sc ii}] than in [O~{\sc ii}]
because of the lower ionization potential of neutral sulfur (10.4\,eV compared
to 13.6\,eV for O$^{\rm o}$). Alternatively, the diffuse emission can be
contaminated by scattered light from the main nebula of higher densities, and
because the [O~{\sc ii}] doublet falls in the near UV wavelength region, the
contamination has a larger effect on [O~{\sc ii}] than on [S~{\sc ii}] (Monk et
al. 1990).

To further explore this problem, we have also carried out a thorough literature
survey of measurements of the [S~{\sc ii}] $\lambda6716$/$\lambda6731$ doublet
ratio amongst ionized gaseous nebulae, emphasizing in particular those expected
to have very low electron densities, such as diffuse ionized gas (DIG),
extragalactic H~{\sc ii} regions and supernova remnants (SNRs). The survey
however does yield some evidence pointing to a $\lambda6716$/$\lambda6731$
ratio in excess of 1.5. Deep spectroscopy of H~{\sc ii} regions in NGC\,55 by
T\"{u}llmann et al. (\cite{tullmann2003}) revealed DIG emission which yields a
[S~{\sc ii}] $\lambda6716/\lambda6731$ ratio of 1.5$\pm0.3$. The error bars
were estimated from their measured flux uncertainties, which were claimed to be
less than 14 per cent.  Galarza, Walterbos \& Braun (\cite{GWB1999}) presented
spectroscopic observations of discrete emission-line nebulae and regions of DIG
in M\,31, including roughly 46 H~{\sc ii} regions and 16 SNRs. In total 15
H~{\sc ii} regions and 4 SNRs have a $\lambda6716/\lambda6731$ ratio larger or
equal to 1.5, with a maximum of $3.9_{2.0}^{5.8}$ (K\,536,an H~{\sc ii}
region).  While the uncertainties were large for most of the individual
measurements, the ratios determined for two H~{\sc ii} regions exceed 1.5 by
more than $2\sigma$, $2.3_{2.2}^{2.4}$ in K\,772 and $2.7_{2.5}^{2.9}$ in K\,70
(Position b; but for ``K\,70 mean'', the ratio is $1.7_{1.6}^{1.8}$).  

The observations of Galarza et al. (\cite{GWB1999}) of H~{\sc ii} regions and
SNRs in M\,31 thus potentially pose a challenge to the validity of {\it
LS}-coupling for the [S~{\sc ii}] doublet.  Given the very low surface
brightness of such emission regions, measurements are always difficult.  Any
new independent measurements to corroborate these results will be invaluable.
Similar observation for the [O~{\sc ii}] $\lambda\lambda$3729, 3726 doublet
ratio should also be attempted, although in this case it is even more difficult
than for the [S~{\sc ii}] lines, given the small wavelength difference between
the two [O~{\sc ii}] lines.

\input{Table06.tex}

Finally, we note that in a recent effort to detect the [$^{13}$C~{\sc iii}]
2s$^2$\,$^1_{1/2}$S$_0$--2s2p\,$^3_{1/2}$P$^{\rm o}_0$\,$\lambda$1910 line,
Rubin et al. (\cite{rubin2004a}) reanalyzed all high resolution spectra of PNe
available from the {\it IUE}\ Final Data Archive. They find that out of 41 PNe
for which the [C~{\sc iii}] 2s$^2$\,$^1$S$_0$--2s2p\,$^3$P$^{\rm
o}_2$\,$\lambda$1907 to C~{\sc iii}] 2s$^2$\,$^1$S$_0$--2s2p\,$^3$P$^{\rm
o}_1$\,$\lambda$1909 ratio is determinable, 8 nebulae, most of them previously
known low density PNe,  show values in excess of its low density limit of 1.51
for pure $^{12}$C and 1.65 for pure $^{13}$C for $T_{\rm e} = 10\ 000$~K.  Not
all of the discrepancies can be explained by measurement uncertainties.  As
they point out, this current theoretical low density limit is fixed simply by
the statistical weights of the energy levels giving rise to the lines, under
the assumption of {\it LS}-coupling. Note that the  [C~{\sc iii}]
2s$^2$\,$^1$S$_0$--2s2p\,$^3$P$^{\rm o}_2$\,$\lambda$1907 line and the C~{\sc
iii}] 2s$^2$\,$^1$S$_0$--2s2p\,$^3$P$^{\rm o}_1$\,$\lambda$1909 line have very
high critical densities of $7.4\times 10^4$ and $9.0\times 10^8$\,cm$^{-3}$.
Thus a PN with a typical electron density of $\la 10\ 000$\,cm$^{-3}$ is
already approaching the low density limit of the $\lambda$1907/$\lambda$1909
ratio. 

In a related but probably of different nature problem, Rubin et al.
(\cite{rubin2004b}) find that for roughly half of the PNe for which the $N_{\rm
e}$-sensitive [Ne~{\sc v}] 14.3/24.3$\mu$m ratio has been determined using {\it
ISO}\ SWS spectra, the observed ratios are out of the low density bound.  Since
these fine-structure lines arise within a single spectral term, i.e.  the
2p$^2$\,$^3$P ground term of [Ne~{\sc v}], the problem here is probably caused
by uncertainties in the relevant collision strengths, rather than related to
the coupling scheme and the validity of the {\it LS}-coupling for these
particular transitions.

\subsubsection{High density limit}
   
As described earlier, at high electron densities, the [O~{\sc ii}]
$\lambda3729/\lambda3726$ doublet ratio is determined entirely by the ratio of
the Einstein transition probabilities, Thus measurements of the doublet ratio
at the high-$N_{\rm e}$ limit can be used to observationally constrain the
$A(^2D_{5/2})/A(^2D_{3/2})$ A ratio. Similar relations hold for the [S~{\sc
ii}], [Cl~{\sc iii}] and [Ar~{\sc iv}] doublet ratios.

In their analysis, Copetti \& Writzl (\cite{copetti2002}) adopted the
transition probabilities for [O~{\sc ii}] tabulated by Wiese et al.
(\cite{wiese}), which yield a very low high-$N_{\rm e}$ limit of 0.26 for the
$\lambda3729/\lambda3726$ doublet ratio. Copetti \& Writzl find that the locus for
the $\lambda3729/\lambda3726$ ratio as a function of the [S~{\sc ii}]
$\lambda6716/\lambda6731$ ratio is incompatible with observations and conclude
that the $A(^2D_{5/2})/A(^2D_{3/2})$ ratio for [O~{\sc ii}] given by Wiese et
al. (\cite{wiese}) needs to be revised upwards by approximately 25--50 per
cent. A similar trend is seen in the current data set -- the long-dashed line
in Fig.\,\ref{f7}, generated using the transition probabilities of Wiese et al.
(\cite{wiese}), deviate significantly from the observations. 

Similar, but to a lesser degree, a discrepancy exists between observations and
the theoretical predictions for the radiative transition probabilities of
[\ion{O}{ii}] given by Zeippen (\cite{zeippen1987a}), and originally calculated
Butler \& Zeippen by (\cite{butler1984}) using the SUPER-STRUCTURE code. In their
calculations, a configuration basis set was selected, which yielded
satisfactory agreement with MC HF-BP (Multi-configuration Hartree-Fock method
with Breit-Pauli relativistic effects) calculations. The effects of
semi-empirical term energy corrections (TEC) and of relativistic
corrections to the magnetic dipole transition operator were fully taken into
account. Loci generated using the [O~{\sc ii}] transition probabilities of
Zeippen (\cite{zeippen1987a}) are shown as short-dashed and
dotted-dotted-dashed lines in Fig.~\ref{f7}. They deviate from the observations
by approximately 20 per cent at the high-$N_{\rm e}$ limit. 

Overall, Fig.\,\ref{f7} shows that amongst the existing atomic parameters for
the [\ion{O}{ii}] forbidden lines, our default set consisting of the transition
probabilities from Zeippen (\cite{zeippen1982}) and collision strengths from
Pradhan (\cite{pradhan}) fit the observations best. Close scrutiny of
Fig.\,\ref{f7} reveals, however, that even this atomic data set shows some
small deviations from the observations at high densities. An upward revision of
the [\ion{O}{ii}] $A(\lambda3729)/A(\lambda3726)$ transition probability ratio
by approximately 6 per cent would improve agreement with observation.

The above discussion is based on the assumption that the transition probabilities
for [S~{\sc ii}] are accurate, and that those of [O~{\sc ii}] are to blame
for any discrepancies shown in Fig.\,\ref{f7} between the observations and
theoretical predictions. Evidence in favour of this interpretation will be
discussed later in the paper.
  
\subsection{[Cl~{\sc iii}] and [Ar~{\sc iv}]}

Cl$^+$ and Cl$^{2+}$ have ionization potentials of 23.8 and 39.6\,eV,
respectively. The corresponding values for Ar$^{2+}$ and Ar$^{3+}$ are 40.7 and
59.8\,eV. Thus [Cl~{\sc iii}] and [Ar~{\sc iv}] emission lines normally arise
from inner regions of higher ionization degrees compared to [S~{\sc ii}] and
[O~{\sc ii}] lines. In addition, the [Cl~{\sc iii}] $\lambda$5517 and
$\lambda$5537 lines have critical densities of $6.4\times 10^3$ and $3.4\times
10^4$\,cm$^{-3}$ at $T_{\rm e} = 10\ 000$~K and the corresponding values for
the [Ar~{\sc iv}] $\lambda$4711 and $\lambda$4740 lines are $1.4\times 10^4$
and $1.3\times 10^5$\,cm$^{-3}$. These values are much higher than those of the
[S~{\sc ii}] and [O~{\sc ii}] doublet lines. In this Section, we will first
compare electron densities derived from the [Cl~{\sc iii}] and [Ar~{\sc iv}]
doublet ratios as well as from the [S~{\sc ii}] and [O~{\sc ii}] doublet
ratios.  This is followed by a brief discussion of the low density limits for
the [Cl~{\sc iii}] and [Ar~{\sc iv}] doublet ratios. Given the weakness of the
[Cl~{\sc iii}] and [Ar~{\sc iv}] lines, the low density limits of their doublet
ratios are much less well constrained by the current observational data set
compared to those for the [S~{\sc ii}] and [O~{\sc ii}] doublet ratios. On the
other hand, given the higher critical densities of the [Cl~{\sc iii}] and
[Ar~{\sc iv}] lines, their low density limits are much easier to achieve than
in the cases of the [S~{\sc ii}] and [O~{\sc ii}] doublet ratios.

\subsubsection{[Cl~{\sc iii}]}

On the whole, densities deduced from the [Cl~{\sc iii}] doublet ratios are
similar to those deduced from the [\ion{O}{ii}] and [\ion{S}{ii}] doublet
ratios.  The agreement between the densities derived from the [\ion{S}{ii}] 
ratio
and from the [\ion{Cl}{iii}] ratio is particularly good (c.f. Figs.~\ref{f11}).  At densities higher
than $\ga 6\times10^3$ cm$^{-3}$, there is evidence that the [\ion{O}{ii}]
doublet ratio yields densities slightly lower than those given by the
[\ion{S}{ii}] ratio and by the [Cl~{\sc iii}] ratio, when using our default
atomic data set (c.f. Figs.~\ref{f8}, \ref{f9}). Note that S$^0$ and S$^+$
have ionization potentials of 10.4 and 23.3\,eV, respectively, compared to
the corresponding values of 13.6 and 35.1\,eV for O$^0$ and O$^+$, and 23.8 and
39.6\,eV for Cl$^+$ and Cl$^{2+}$, respectively.  Thus the O$^+$ ionization
zone is expected to have a larger overlap with that of Cl$^{2+}$. This fact,
combined with the realization that the [\ion{O}{ii}] and [\ion{S}{ii}] doublets
have comparable critical densities which are smaller than those for the
[\ion{Cl}{iii}] doublet, suggests that the systematically lower densities found
from the [\ion{O}{ii}] doublet compared to those found from the [\ion{S}{ii}]
and [\ion{Cl}{iii}] doublets at densities higher than $\ga 6\times10^3$
cm$^{-3}$ is unlikely to be caused by density inhomogeneities or ionization
stratification. The discrepancy is more likely caused by uncertainties in the
transition probabilities for the [\ion{O}{ii}] forbidden lines; as estimated in
the previous section, the [\ion{O}{ii}] $A(\lambda3729)/A(\lambda3726)$ ratio
given by Zeippen (\cite{zeippen1982}) may need to be revised upwards by
$\approx$ 6 per cent.

\subsubsection{[Ar~{\sc iv}]}

Figs.~\ref{f8}- \ref{f13}, where we compare densities derived from the four
optical $N_{\rm e}$ diagnostics, are arranged in such a way that for each pair
of diagnostic ratios, $N_{\rm e}$ derived from the diagnostic of lower critical
density is used as the abscissa.  The [S~{\sc ii}] and [O~{\sc ii}] doublets
have comparable critical densities; here we have chosen the [O~{\sc ii}]
$N_{\rm e}$ as the abscissa. 

Plotted in this way, Figs.~\ref{f10}, \ref{f12} and \ref{f13}, where densities
derived from [Ar~{\sc iv}] are compared to those resulting from the other three
diagnostic ratios, show a common feature -- a rise in the data points as the high
$N_{\rm e}$ limit is approached, suggesting densities derived from the [Ar~{\sc
iv}] doublet are systematically higher than those derived from the other three
$N_{\rm e}$ indicators, and that the deviation increases with increasing
$N_{\rm e}$.  Adopting the earlier collision strengths of Zeippen,
Butler \& Le Bourlot (\cite{zeippen1987b}) would lead to even higher densities
being determined from the [Ar~{\sc iv}] doublet, although the increase is less than
0.1\,dex except for densities lower than 10$^3$ cm$^{-3}$ (Keenan et al.
(\cite{keenan1997}).  At such low densities, the [Ar~{\sc iv}] doublet ratio is
however no longer a sensitive $N_{\rm e}$-diagnostic. 

Using the Long Wavelength Spectrometer ({\it LWS}) on board the Infrared Space
Observatory ({\it ISO}), Liu et al. (\cite{Liu2001}) determined electron
densities for a sample of bright Galactic PNe from the [O~{\sc iii}]
52$\mu$m/88$\mu$m fine-structure line ratio and found that they are
systematically lower than those derived from the optical [Cl~{\sc iii}] and
[Ar~{\sc iv}] doublet ratios. Considering the much lower critical densities of
these [O~{\sc iii}] lines ($\sim$ a few $\times 10^3$\,cm$^{-3}$) compared to
those of the [Cl~{\sc iii}] and [Ar~{\sc iv}] optical lines and the fact that
the [Cl~{\sc iii}] and [Ar~{\sc iv}] ratios yielded comparable densities
consistent with each other, they argued that the discrepancy is unlikely to be
caused by ionization stratification. Liu et al attributed it to the presence of
modest $N_{\rm e}$ inhomogeneities. Emission of the [O~{\sc iii}]
fine-structure lines is suppressed by collisional de-excitation by electron
impacts in high $N_{\rm e}$ regions. Our current optical study for a much
larger sample of PNe however shows some evidence that the [Ar~{\sc iv}] doublet
ratio tends to yield densities systematically higher than those deduced from
the [Cl~{\sc iii}] (and [O~{\sc ii}] or [S~{\sc ii}]) doublet ratio. The
deviation is most profound at densities $\ga 10^4$\,cm$^{-3}$ and increases as
$N_{\rm e}$ increases (c.f.  Fig.\,\ref{f12}).  Since Ar$^{2+}$ has an
ionization potential of 40.7\,eV, much higher than 23.8\,eV of Cl$^+$, it is
possible that part of the discrepancy between the [Cl~{\sc iii}] and [Ar~{\sc
iv}] densities shown in Fig.\,\ref{f13} is caused by ionization stratification,
where the inner, higher ionization Ar$^{3+}$ zone has a higher $N_{\rm e}$ than
the outer, lower ionization Cl$^{2+}$ zone. Alternatively, nebulae with
densities exceeding $\sim 10^4$\,cm$^{-3}$ may contain ionized condensations
with $N_{\rm e}$ exceeding the critical density for [Cl~{\sc iii}], [O~{\sc
ii}] and [S~{\sc ii}], with the [Ar~{\sc iv}] emission is less suppressed by
collisional de-excitation.
   
\subsubsection{Low density limit}

The [Cl~{\sc iii}] $\lambda5517/\lambda5537$ and [Ar~{\sc iv}]
$\lambda4711/\lambda4740$ doublet ratios have low density limits of 1.44 and
1.42, respectively at $T_{\rm e} = 10\ 000$~K.  For the [Cl~{\sc iii}], only
two nebulae in our sample have $\lambda5517/\lambda5537$ measured ratios in
excess of its low density limit, $1.51\pm0.28$ in NGC\,6072 and $1.46\pm0.22$
in NGC\,5307. In both cases, the results remain consistent with the low density
limit within 1$\sigma$ measurement uncertainty. As described in \S{2} the [Cl
~{\sc iii}] doublet lines from NGC~6072 are very weak, and the high
$\lambda5517/\lambda5537$ ratio measured for this nebula is most likely caused
by observational uncertainties.

Similarly, two PNe in our sample have their measured [Ar~{\sc iv}]
$\lambda4711/\lambda4740$ doublet ratios in excess of the low density limit,
$2.39\pm$0.65 in NGC\,6781 and $1.64\pm 0.12$ in NGC\,6072.  In both cases, the
uncertainties are large. Accurate measurements of the $\lambda4711/\lambda4740$
doublet ratio in these nebulae were further hampered by the serious
contamination of the [Ar~{\sc iv}] $\lambda4711$ line by the nearby strong He~I
$\lambda4713$ line. 

To conclude, amongst the PNe in our current sample, no credible cases can be
established that the [Cl~{\sc iii}] $\lambda5517/\lambda5537$ and [Ar~{\sc iv}]
$\lambda4711/\lambda4740$ doublet ratios have a measured value in excess of
their low density limit.  Further high signal-to-noise ratio and high spectral
resolution measurements of these two doublet ratios in low density nebulae
should be useful. 

Finally we note that Peimbert (\cite{peimbert04}) has recently obtained deep
echelle spectrum of the Giant H~{\sc ii} region 30 Doradus using the
Ultraviolet Visual Echelle Spectrograph (UVES) mounted on the VLT Kueyen
Telescope, at a spectral resolution of $\lambda/\Delta\lambda\sim 8800$.  The
region he observed is found to have a very low $N_{\rm e}$ of just a few
hundred cm$^{-3}$. The spectrum yields a [Cl~{\sc iii}]
$\lambda5517/\lambda5537$ and [Ar~{\sc iv}] $\lambda4711/\lambda4740$ doublet
ratio of 1.33 and 1.35, respectively. Again both values are well within the
low-$N_{\rm e}$ limits of the two doublet ratios.

\subsection{Kolmogorov-Smirnov test}

In order to investigate whether the four optical $N_{\rm e}$-diagnostics yield
electron densities compatible to each other, the Kolmogorov-Smirnov Two-sample Test
has been carried out on the densities derived from each pair of diagnostic ratio
,with the Null Hypothesis that densities derived from the two diagnostics of the
pair are drawn from the same sample.  Results from the tests are listed in
Table~\ref{kstest}. The table shows that except for the case of densities derived 
from [O~{\sc ii}] versus those deduced from [Ar~{\sc iv}], the Null Hypothesis cannot be
rejected at a high confidence level.  In the case of [O~{\sc ii}] versus
[Ar~{\sc iv}] the Hypothesis can be rejected at a confidence level of 90 per
cent, i.e. there is good evidence suggesting densities derived from the [O~{\sc
ii}] ratio differ systematically from those yielded by the [Ar~{\sc iv}] ratio. There
is also some evidence, albeit quite weak, that densities yielded by the
[S~{\sc ii}] and [Cl~{\sc iii}] ratios differ from those derived from the
[Ar~{\sc iv}] doublet -- the Null Hypothesis can be rejected with a risk
possibility no more than 51 per cent.  The evidence becomes more apparent
if we restrict our analysis to high $N_{\rm e}$ nebulae. In fact, visual 
inspection of Figs.~\ref{f8}- \ref{f13} suggests that in general, 
\[
N_{\rm e}([\ion{O}{ii}]) \la N_{\rm e}([\ion{S}{ii}]) \approx N_{\rm e}([\ion{Cl}{iii}])<
 N_{\rm e}([\ion{Ar}{iv}]),
\]
and the differences between the densities deduced from the [Ar~{\sc iv}]
doublet ratio and those derived from [S~{\sc ii}] or [Cl~{\sc iii}] increase 
as $N_{\rm e}$ increases.
\input{Table07.tex}

\section{Summary}
 
In this paper we have presented measurements of the four optical $N_{\rm e}$-diagnostic
ratios for a sample of $>100$ Galactic PNe. We show that by adopting
the default atomic parameters listed in Table~\ref{atomic_ref}, we obtain densities 
from the four diagnostics that are in general good agreement. The agreement between
the [S~{\sc ii}] and [Cl~{\sc iii}] densities is particularly good. We show
that, under the assumption that the [O~{\sc ii}] and [S~{\sc ii}] lines
arise from similar ionization zones (and would then yield comparable
densities), the transition probabilities of Zeippen (\cite{zeippen1982})
combined with the collision strengths of Pradhan (\cite{pradhan}) for O$^+$ 
give the best fit to observations.  The more recent calculations of
[\ion{O}{ii}] transition probabilities by Zeippen (\cite{zeippen1987a}) give a
poor fit to the measurements at high densities and therefore can be ruled out.  We
confirm the earlier result of Copetti \& Writzl (\cite{copetti2002}) that when
using the [\ion{O}{ii}] transition probabilities listed in Wiese et al.
(\cite{wiese}), electron densities derived from the [\ion{O}{ii}] doublet ratio
are systematically lower than those deduced from the [\ion{S}{ii}] lines and
that the discrepancies are most likely caused by uncertainties in the
[O~{\sc ii}] transition probabilities of Wiese et al.

At high electron densities, the [O~{\sc ii}] doublet ratio is found to yield
slightly lower densities compared to those derived from the [S~{\sc ii}] and
[Cl~{\sc iii}] lines. The discrepancy can be removed by increasing the [O~{\sc
ii}] $\lambda$3729 to $\lambda$3726 transition probability ratio of Zeippen
(\cite{zeippen1982}) by approximately 6 per cent.

We also find that at high nebular densities, electron densities derived from the [Ar~{\sc iv}]
ratio are systematically higher than deduced from the other three diagnostic
ratios, and that the deviations increase with increasing density. While it is
difficult to rule out that this is not caused by uncertainties in the [Ar~{\sc
iv}] transition probabilities, modest density inhomogeneities, as pointed out
by Liu et al. (\cite{Liu2001}), are probably to blame for the deviations.

Amongst the large number of PNe observed here as well as PNe and H~{\sc ii}
regions studied in the literature, no credible measurements, when measurement
uncertainties are taken into account, can be established that yield an [O~{\sc
ii}] $\lambda3729/\lambda3726$ doublet ratio that exceeds the standard
low-$N_{\rm e}$ limit of 1.5 under the assumption of {\it LS}-coupling.  Thus
we find no evidence in support of the more recent calculations of the [O~{\sc
ii}] collision strengths by McLaughlin \& Bell (\cite{mclaughlin}) who found a
much higher low density limit of 1.93. Their result was attributed to
departures from pure {\it LS}-coupling owing to relativistic effects.  On the
other hand, a thorough literature survey reveals a few measurements of the
[S~{\sc ii}] doublet in diffuse ionized gas in extragalactic H~{\sc ii} regions
and SNRs that produce a [S~{\sc ii}] doublet ratio in excess of its low-$N_{\rm
e}$ limit of 1.5 by more than a 2$\sigma$ measurement uncertainty.  Further
observations to verify these results will be valuable.

\begin{acknowledgements} 
We are grateful to Y. Liu and S.-G. Luo for their help with the preparation of
this paper and thank P.J. Storey for discussion. We would also like to thank
Dr. R. Rubin for a critical reading of the manuscript prior to its submission.
\end{acknowledgements}

\end{document}

%% file: Table01.tex
\begin{table}
\caption{Spectral wavelength coverage and resolution of the spectra}
\label{jan}
\centering
\begin{tabular}{cccccc}
\hline
\hline
\noalign{\smallskip}
Date &Telescope& $\lambda$-range &Slit width& FWHM \\
     &         & ({\AA}) &(arcsec)& ({\AA}) \\
\noalign{\smallskip}
\hline
\noalign{\smallskip}
1995 Jul&ESO 1.52m & 3530--7430&2&4.5\\
1995 Jul&ESO 1.52m & 4000--4987&2&1.5\\
1996 Jul&ESO 1.52m & 3530--7430&2&4.5\\
1996 Jul&ESO 1.52m & 4000--4987&2&1.5\\
2001 Jun&ESO 1.52m & 3500--4805&2&1.5\\
2001 Aug&INT 2.5m   & 3610--4410&1&1.5\\
\noalign{\smallskip}
\hline
\end{tabular}
\end{table}

%% file: Table02.tex
\setcounter{table}{1}
 \begin{table*}
\caption{\label{source} Line ratios and electron densities derived using the default atomic parameters described in Table~3}
\centering
\begin{tabular}{l c c c c c c c c c c}
\hline
\hline
\noalign{\smallskip}
       & \multicolumn{2}{c}{[O~{\sc ii}]}  & \multicolumn{2}{c}{[S~{\sc ii}]} & \multicolumn{2}{c}{[Cl~{\sc iii}]} & \multicolumn{2}{c}{[Ar~{\sc iv}]}\\
Source$^a$ & $\lambda3729/$ & ${\rm Log} N_{\rm e}$ & $\lambda6716/$ & ${\rm Log} N_{\rm e}$ &$\lambda5517/$ & ${\rm Log} N_{\rm e}$ & $\lambda4711/$ & ${\rm Log} N_{\rm e}$\\
&$\lambda3726$ & (cm$^{-3}$) &$\lambda6731$ & (cm$^{-3}$) &  $\lambda5537$ & (cm$^{-3}$) & $\lambda4740$ & (cm$^{-3}$)&\\
\noalign{\smallskip}
 \hline
 \noalign{\smallskip}
  BoBn1    &              &                                                                 &                 &                                                                  &                 &                                                                &    1.712$\pm.422$&  $ < $1.26\\  
  Cn1-5$^s$&              &                                                                 &  0.585 $\pm.010$&  3.68$\stackrel{\scriptscriptstyle+.03}{\scriptscriptstyle-.03 }$& 0.826 $\pm.041$&  3.53$\stackrel{\scriptscriptstyle+.05}{\scriptscriptstyle-.06 }$&   1.159$\pm.106$&  3.15$\stackrel{\scriptscriptstyle+.20}{\scriptscriptstyle-.36 }$ \\
  Cn2-1    &.480$\pm.010$&  3.71$\stackrel{\scriptscriptstyle+.04}{\scriptscriptstyle-.04 }$&  0.549 $\pm.021$&  3.81$\stackrel{\scriptscriptstyle+.10}{\scriptscriptstyle-.09 }$& 0.602 $\pm.069$&  3.87$\stackrel{\scriptscriptstyle+.10}{\scriptscriptstyle-.12 }$&   0.472$\pm.016$&  4.23$\stackrel{\scriptscriptstyle+.03}{\scriptscriptstyle-.03 }$ \\
  Cn3-1    &.487$\pm.005$&  3.69$\stackrel{\scriptscriptstyle+.02}{\scriptscriptstyle-.02 }$&                 &                                                                  &                &                                                                  &                    &                                                                   \\ 
  H1-35    &.378$\pm.017$&  4.37$\stackrel{\scriptscriptstyle+.32}{\scriptscriptstyle-.18 }$&  0.483 $\pm.033$&  4.29$\stackrel{\scriptscriptstyle+1.6}{\scriptscriptstyle-.29 }$& 0.429 $\pm.074$&  4.20$\stackrel{\scriptscriptstyle+.17}{\scriptscriptstyle-.19 }$&   1.302$\pm.256$&  2.67$\stackrel{\scriptscriptstyle+.63}{\scriptscriptstyle-2.7 }$ \\
  H1-36    &             &                                                                  &  0.503 $\pm.023$&  4.09$\stackrel{\scriptscriptstyle+.21}{\scriptscriptstyle-.16 }$& 0.397 $\pm.039$&  4.28$\stackrel{\scriptscriptstyle+.11}{\scriptscriptstyle-.11 }$&   0.253$\pm.008$&  4.74$\stackrel{\scriptscriptstyle+.03}{\scriptscriptstyle-.03 }$ \\
  H1-41    &             &                                                                  &  0.813 $\pm.040$&  3.12$\stackrel{\scriptscriptstyle+.08}{\scriptscriptstyle-.08 }$& 1.101 $\pm.150$&  3.12$\stackrel{\scriptscriptstyle+.21}{\scriptscriptstyle-.39 }$&   1.214$\pm.063$&  3.01$\stackrel{\scriptscriptstyle+.16}{\scriptscriptstyle-.23 }$ \\
  H1-42    &             &                                                                  &  0.532 $\pm.014$&  3.91$\stackrel{\scriptscriptstyle+.09}{\scriptscriptstyle-.08 }$& 0.719 $\pm.093$&  3.69$\stackrel{\scriptscriptstyle+.12}{\scriptscriptstyle-.15 }$&   0.704$\pm.020$&  3.88$\stackrel{\scriptscriptstyle+.03}{\scriptscriptstyle-.03 }$ \\
  H1-50    &.455$\pm.016$&  3.81$\stackrel{\scriptscriptstyle+.08}{\scriptscriptstyle-.07 }$&  0.538 $\pm.017$&  3.88$\stackrel{\scriptscriptstyle+.09}{\scriptscriptstyle-.08 }$& 0.585 $\pm.058$&  3.89$\stackrel{\scriptscriptstyle+.09}{\scriptscriptstyle-.11 }$&   0.565$\pm.013$&  4.08$\stackrel{\scriptscriptstyle+.02}{\scriptscriptstyle-.02 }$ \\
  H1-54    &             &                                                                  &  0.493 $\pm.017$&  4.18$\stackrel{\scriptscriptstyle+.18}{\scriptscriptstyle-.14 }$& 0.521 $\pm.079$&  4.01$\stackrel{\scriptscriptstyle+.14}{\scriptscriptstyle-.16 }$&                 &                                                                   \\
  He2-113  &.492$\pm.027$&  3.67$\stackrel{\scriptscriptstyle+.10}{\scriptscriptstyle-.09 }$&                 &                                                                  &                &                                                                  &                 &                                                                   \\
  He2-118  &.440$\pm.012$&  3.89$\stackrel{\scriptscriptstyle+.06}{\scriptscriptstyle-.06 }$&  0.518 $\pm.032$&  3.99$\stackrel{\scriptscriptstyle+.24}{\scriptscriptstyle-.18 }$& 0.541 $\pm.082$&  3.97$\stackrel{\scriptscriptstyle+.14}{\scriptscriptstyle-.16 }$&   0.408$\pm.017$&  4.34$\stackrel{\scriptscriptstyle+.03}{\scriptscriptstyle-.03 }$\\
  He2-136  &             &                                                                  &  0.578 $\pm.017$&  3.69$\stackrel{\scriptscriptstyle+.06}{\scriptscriptstyle-.06 }$& 0.855 $\pm.131$&  3.49$\stackrel{\scriptscriptstyle+.17}{\scriptscriptstyle-.24 }$&   1.040$\pm.033$&  3.40$\stackrel{\scriptscriptstyle+.05}{\scriptscriptstyle-.06 }$\\
  He2-142  &             &                                                                  &  0.424 $\pm.022$&  3.63$\stackrel{\scriptscriptstyle+.00}{\scriptscriptstyle-.00 }$& 0.433 $\pm.073$&  4.19$\stackrel{\scriptscriptstyle+.17}{\scriptscriptstyle-.19 }$&                 &                                                                  \\
  He2-185  &             &                                                                  &  0.617 $\pm.030$&  3.56$\stackrel{\scriptscriptstyle+.09}{\scriptscriptstyle-.09 }$& 0.893 $\pm.080$&  3.44$\stackrel{\scriptscriptstyle+.11}{\scriptscriptstyle-.13 }$&   0.781$\pm.024$&  3.78$\stackrel{\scriptscriptstyle+.03}{\scriptscriptstyle-.03 }$\\
  He2-434  &             &                                                                  &  0.571 $\pm.078$&  3.71$\stackrel{\scriptscriptstyle+.36}{\scriptscriptstyle-.28 }$& 1.000 $\pm.150$&  3.29$\stackrel{\scriptscriptstyle+.19}{\scriptscriptstyle-.31 }$&   0.926$\pm.026$&  3.58$\stackrel{\scriptscriptstyle+.04}{\scriptscriptstyle-.05 }$\\
  He2-90   &             &                                                                  &  0.439 $\pm.042$&  3.44$\stackrel{\scriptscriptstyle+.00}{\scriptscriptstyle-.79 }$& 0.752 $\pm.130$&  3.64$\stackrel{\scriptscriptstyle+.16}{\scriptscriptstyle-.23 }$&                 &                                                                  \\
  He2-97   &             &                                                                  &  0.495 $\pm.027$&  4.17$\stackrel{\scriptscriptstyle+.34}{\scriptscriptstyle-.22 }$& 0.345 $\pm.040$&  4.45$\stackrel{\scriptscriptstyle+.15}{\scriptscriptstyle-.14 }$&   0.588$\pm.055$&  4.05$\stackrel{\scriptscriptstyle+.08}{\scriptscriptstyle-.09 }$\\
  He3-1333 &.916$\pm.030$&  2.84$\stackrel{\scriptscriptstyle+.05}{\scriptscriptstyle-.05 }$&                 &                                                                  &                &                                                                  &                 &                                                                  \\
  Hu1-2    &.487$\pm.011$&  3.69$\stackrel{\scriptscriptstyle+.04}{\scriptscriptstyle-.04 }$&  0.595 $\pm.014$&  3.64$\stackrel{\scriptscriptstyle+.05}{\scriptscriptstyle-.05 }$& 0.787 $\pm.031$&  3.59$\stackrel{\scriptscriptstyle+.05}{\scriptscriptstyle-.05 }$&   0.971$\pm.019$&  3.50$\stackrel{\scriptscriptstyle+.03}{\scriptscriptstyle-.03 }$\\
  Hu2-1    &.424$\pm.006$&  3.97$\stackrel{\scriptscriptstyle+.04}{\scriptscriptstyle-.04 }$&                 &                                                                  &                &                                                                  &                 &                                                                  \\
  IC1297   &.570$\pm.022$&  3.45$\stackrel{\scriptscriptstyle+.05}{\scriptscriptstyle-.05 }$&  0.645 $\pm.021$&  3.48$\stackrel{\scriptscriptstyle+.05}{\scriptscriptstyle-.05 }$& 0.943 $\pm.071$&  3.36$\stackrel{\scriptscriptstyle+.09}{\scriptscriptstyle-.11 }$&   1.009$\pm.021$&  3.44$\stackrel{\scriptscriptstyle+.04}{\scriptscriptstyle-.03 }$\\
  IC3568   &.626$\pm.025$&  3.33$\stackrel{\scriptscriptstyle+.05}{\scriptscriptstyle-.05 }$&  0.775 $\pm.102$&  3.19$\stackrel{\scriptscriptstyle+.19}{\scriptscriptstyle-.22 }$& 1.427 $\pm.079$&  1.59$\stackrel{\scriptscriptstyle+.83}{\scriptscriptstyle-.00 }$&   1.145$\pm.028$&  3.18$\stackrel{\scriptscriptstyle+.06}{\scriptscriptstyle-.06 }$\\
 IC4191$^s$&             &                                                                  &  0.535 $\pm.014$&  3.89$\stackrel{\scriptscriptstyle+.08}{\scriptscriptstyle-.07 }$& 0.541 $\pm.032$&  3.97$\stackrel{\scriptscriptstyle+.06}{\scriptscriptstyle-.06 }$&   0.588$\pm.014$&  4.05$\stackrel{\scriptscriptstyle+.02}{\scriptscriptstyle-.02 }$\\
 IC4406$^s$&             &                                                                  &  1.027 $\pm.030$&  2.73$\stackrel{\scriptscriptstyle+.05}{\scriptscriptstyle-.05 }$& 1.065 $\pm.092$&  3.18$\stackrel{\scriptscriptstyle+.13}{\scriptscriptstyle-.19 }$&   2.000$\pm.204$&  $<$1.26                                                          \\
  IC4593   &             &                                                                  &  0.637 $\pm.073$&  3.51$\stackrel{\scriptscriptstyle+.22}{\scriptscriptstyle-.20 }$& 0.667 $\pm.107$&  3.77$\stackrel{\scriptscriptstyle+.14}{\scriptscriptstyle-.18 }$&                 &                                                                  \\
  IC4634   &.438$\pm.019$&  3.89$\stackrel{\scriptscriptstyle+.11}{\scriptscriptstyle-.09 }$&  0.521 $\pm.014$&  3.97$\stackrel{\scriptscriptstyle+.10}{\scriptscriptstyle-.08 }$& 0.699 $\pm.068$&  3.72$\stackrel{\scriptscriptstyle+.10}{\scriptscriptstyle-.11 }$&   0.735$\pm.016$&  3.84$\stackrel{\scriptscriptstyle+.02}{\scriptscriptstyle-.02 }$\\
  IC4637   &             &                                                                  &                 &                                                                  &                &                                                                  &   0.746$\pm.117$&  3.83$\stackrel{\scriptscriptstyle+.14}{\scriptscriptstyle-.20 }$\\
  IC4699   &             &                                                                  &  0.649 $\pm.089$&  3.47$\stackrel{\scriptscriptstyle+.25}{\scriptscriptstyle-.24 }$& 1.359 $\pm.201$&  2.38$\stackrel{\scriptscriptstyle+.57}{\scriptscriptstyle-2.3 }$&   1.261$\pm.029$&  2.86$\stackrel{\scriptscriptstyle+.09}{\scriptscriptstyle-.14 }$\\
  IC4776   &.437$\pm.018$&  3.90$\stackrel{\scriptscriptstyle+.10}{\scriptscriptstyle-.08 }$&  0.505 $\pm.023$&  4.08$\stackrel{\scriptscriptstyle+.21}{\scriptscriptstyle-.16 }$& 0.565 $\pm.035$&  3.93$\stackrel{\scriptscriptstyle+.06}{\scriptscriptstyle-.06 }$&   0.340$\pm.023$&  4.49$\stackrel{\scriptscriptstyle+.05}{\scriptscriptstyle-.06 }$\\
  IC4846   &             &                                                                  &  0.546 $\pm.018$&  3.83$\stackrel{\scriptscriptstyle+.09}{\scriptscriptstyle-.08 }$& 0.758 $\pm.086$&  3.63$\stackrel{\scriptscriptstyle+.11}{\scriptscriptstyle-.14 }$&   0.746$\pm.028$&  3.83$\stackrel{\scriptscriptstyle+.04}{\scriptscriptstyle-.04 }$\\
  IC4997   &.393$\pm.016$&  4.20$\stackrel{\scriptscriptstyle+.19}{\scriptscriptstyle-.13 }$&  0.485 $\pm.009$&  4.24$\stackrel{\scriptscriptstyle+.12}{\scriptscriptstyle-.10 }$& 0.508 $\pm.057$&  4.03$\stackrel{\scriptscriptstyle+.11}{\scriptscriptstyle-.12 }$&   0.280$\pm.026$&  4.65$\stackrel{\scriptscriptstyle+.08}{\scriptscriptstyle-.08 }$\\
  IC5217   &.456$\pm.013$&  3.81$\stackrel{\scriptscriptstyle+.06}{\scriptscriptstyle-.05 }$&                 &                                                                  &                &                                                                  &                 &                                                                  \\
  M1-20    &.420$\pm.018$&  4.00$\stackrel{\scriptscriptstyle+.12}{\scriptscriptstyle-.10 }$&  0.524 $\pm.016$&  3.95$\stackrel{\scriptscriptstyle+.11}{\scriptscriptstyle-.10 }$& 0.775 $\pm.126$&  3.60$\stackrel{\scriptscriptstyle+.16}{\scriptscriptstyle-.21 }$&   0.599$\pm.025$&  4.03$\stackrel{\scriptscriptstyle+.04}{\scriptscriptstyle-.04 }$\\
  M1-29    &             &                                                                  &  0.629 $\pm.008$&  3.53$\stackrel{\scriptscriptstyle+.03}{\scriptscriptstyle-.03 }$& 0.800 $\pm.115$&  3.57$\stackrel{\scriptscriptstyle+.14}{\scriptscriptstyle-.19 }$&                 &                                                                  \\
  M1-42    &.708$\pm.051$&  3.12$\stackrel{\scriptscriptstyle+.09}{\scriptscriptstyle-.09 }$&  0.826 $\pm.034$&  3.09$\stackrel{\scriptscriptstyle+.07}{\scriptscriptstyle-.07 }$& 1.050 $\pm.138$&  3.23$\stackrel{\scriptscriptstyle+.18}{\scriptscriptstyle-.29 }$&   1.295$\pm.045$&  2.69$\stackrel{\scriptscriptstyle+.19}{\scriptscriptstyle-.35 }$\\
  M1-61    &             &                                                                  &  0.485 $\pm.014$&  4.26$\stackrel{\scriptscriptstyle+.18}{\scriptscriptstyle-.14 }$& 0.437 $\pm.050$&  4.18$\stackrel{\scriptscriptstyle+.12}{\scriptscriptstyle-.12 }$&   0.394$\pm.025$&  4.37$\stackrel{\scriptscriptstyle+.05}{\scriptscriptstyle-.05 }$\\
  M1-64    &1.23$\pm0.01$&  2.32$\stackrel{\scriptscriptstyle+.03}{\scriptscriptstyle-.03 }$&                 &                                                                  &                &                                                                  &                 &                                                                   \\
  M2-23    &             &                                                                  &  0.483 $\pm.047$&  4.29$\stackrel{\scriptscriptstyle+.00}{\scriptscriptstyle-.40 }$& 0.599 $\pm.093$&  3.87$\stackrel{\scriptscriptstyle+.14}{\scriptscriptstyle-.17 }$&   0.263$\pm.016$&  4.70$\stackrel{\scriptscriptstyle+.05}{\scriptscriptstyle-.05 }$\\
  M2-24    &.424$\pm.036$&  3.99$\stackrel{\scriptscriptstyle+.09}{\scriptscriptstyle-.08 }$&                 &                                                                  &                &                                                                  &   0.242$\pm.043$&  4.78$\stackrel{\scriptscriptstyle+.16}{\scriptscriptstyle-.17 }$\\
  M2-27    &             &                                                                  &  0.535 $\pm.011$&  3.88$\stackrel{\scriptscriptstyle+.07}{\scriptscriptstyle-.06 }$& 0.461 $\pm.045$&  4.12$\stackrel{\scriptscriptstyle+.09}{\scriptscriptstyle-.10 }$&                 &                                                                  \\
  M2-31    &             &                                                                  &  0.552 $\pm.018$&  3.80$\stackrel{\scriptscriptstyle+.08}{\scriptscriptstyle-.07 }$& 0.606 $\pm.055$&  3.86$\stackrel{\scriptscriptstyle+.08}{\scriptscriptstyle-.09 }$&                 &                                                                  \\
  M2-33    &             &                                                                  &  0.746 $\pm.056$&  3.25$\stackrel{\scriptscriptstyle+.12}{\scriptscriptstyle-.13 }$&                &                                                                  &   1.148$\pm.141$&  3.18$\stackrel{\scriptscriptstyle+.24}{\scriptscriptstyle-.56 }$\\
\noalign{\smallskip}                                                                                                                                                                                                                                                                 
  \hline                                                                                                                                                                                                                                                                              
  \end{tabular}                                                                                                                                                                                                                                                                       
\end{table*}                                                                                                                                                                                                                                                                         
                                                                                                                                                                                                                                                                                     
\setcounter{table}{1}                                                                                                                                                                                                                                                                
\begin{table*}                                                                                                                                                                                                                                                                       
\caption{ {  \it --continued} }                                                                                                                                                                                                                                                      
\centering                                                                                                                                                                                                                                                                           
\begin{tabular}{l c c c c c c c c c c}                                                                                                                                                                                                                                               
\hline                                                                                                                                                                                                                                                                               
\hline                                                                                                                                                                                                                                                                               
\noalign{\smallskip}                                                                                                                                                                                                                                                                 
       & \multicolumn{2}{c}{[O~{\sc ii}]}  & \multicolumn{2}{c}{[S~{\sc ii}]} & \multicolumn{2}{c}{[Cl~{\sc iii}]} & \multicolumn{2}{c}{[Ar~{\sc iv}]}\\                                                                                                                             
Source$^a$ & $\lambda3729/$ & ${\rm Log} N_{\rm e}$ & $\lambda6716/$ & ${\rm Log} N_{\rm e}$ &$\lambda5517/$ & ${\rm Log} N_{\rm e}$ & $\lambda4711/$ & ${\rm Log} N_{\rm e}$\\                                                                                                          
&$\lambda3726$ & (cm$^{-3}$) &$\lambda6731$ & (cm$^{-3}$) &  $\lambda5537$ & (cm$^{-3}$) & $\lambda4740$ & (cm$^{-3}$)&\\                                                                                                                                                            
\noalign{\smallskip}                                                                                                                                                                                                                                                                 
\hline                                                                                                                                                                                                                                                                               
                                                                                                                                                                                                                                                                                     
  M2-36    &.509$\pm.035$&  3.62$\stackrel{\scriptscriptstyle+.12}{\scriptscriptstyle-.10 }$& 0.599  $\pm.032$&  3.62$\stackrel{\scriptscriptstyle+.11}{\scriptscriptstyle-.11 }$& 0.709 $\pm.030$&  3.70$\stackrel{\scriptscriptstyle+.04}{\scriptscriptstyle-.05 }$&   0.870$\pm.023$&  3.65$\stackrel{\scriptscriptstyle+.04}{\scriptscriptstyle-.03 }$\\
  M2-39    &             &                                                                  & 0.578  $\pm.023$&  3.69$\stackrel{\scriptscriptstyle+.10}{\scriptscriptstyle-.09 }$&                &                                                                  &   0.800$\pm.115$&  3.75$\stackrel{\scriptscriptstyle+.14}{\scriptscriptstyle-.19 }$\\
  M2-4     &.482$\pm.014$&  3.71$\stackrel{\scriptscriptstyle+.05}{\scriptscriptstyle-.05 }$& 0.549  $\pm.024$&  3.81$\stackrel{\scriptscriptstyle+.12}{\scriptscriptstyle-.11 }$& 0.641 $\pm.058$&  3.80$\stackrel{\scriptscriptstyle+.08}{\scriptscriptstyle-.09 }$&   0.610$\pm.048$&  4.02$\stackrel{\scriptscriptstyle+.06}{\scriptscriptstyle-.07 }$\\    
  M2-42    &             &                                                                  & 0.621  $\pm.019$&  3.55$\stackrel{\scriptscriptstyle+.06}{\scriptscriptstyle-.06 }$& 0.901 $\pm.162$&  3.42$\stackrel{\scriptscriptstyle+.20}{\scriptscriptstyle-.31 }$&                 &                                                                  \\
  M2-6     &             &                                                                  & 0.532  $\pm.028$&  3.91$\stackrel{\scriptscriptstyle+.17}{\scriptscriptstyle-.14 }$&                &                                                                  &   1.168$\pm.203$&  3.14$\stackrel{\scriptscriptstyle+.33}{\scriptscriptstyle-1.9 }$\\
  M3-21    &.465$\pm.008$&  3.78$\stackrel{\scriptscriptstyle+.03}{\scriptscriptstyle-.03 }$& 0.503  $\pm.018$&  4.10$\stackrel{\scriptscriptstyle+.15}{\scriptscriptstyle-.12 }$& 0.515 $\pm.043$&  4.02$\stackrel{\scriptscriptstyle+.08}{\scriptscriptstyle-.09 }$&   0.395$\pm.008$&  4.34$\stackrel{\scriptscriptstyle+.02}{\scriptscriptstyle-.02 }$\\
  M3-27    &.584$\pm.024$&  3.42$\stackrel{\scriptscriptstyle+.06}{\scriptscriptstyle-.05 }$&                 &                                                                  &                &                                                                  &                 &                                                                  \\
  M3-29    &             &                                                                  & 0.917  $\pm.034$&  2.93$\stackrel{\scriptscriptstyle+.06}{\scriptscriptstyle-.06 }$& 1.086 $\pm.252$&  3.15$\stackrel{\scriptscriptstyle+.55}{\scriptscriptstyle-1.8 }$&                 &                                                                   \\
  M3-32    &.518$\pm.015$&  3.59$\stackrel{\scriptscriptstyle+.05}{\scriptscriptstyle-.04 }$& 0.633  $\pm.036$&  3.52$\stackrel{\scriptscriptstyle+.10}{\scriptscriptstyle-.10 }$& 0.893 $\pm.112$&  3.44$\stackrel{\scriptscriptstyle+.14}{\scriptscriptstyle-.19 }$&   1.178$\pm.050$&  3.11$\stackrel{\scriptscriptstyle+.11}{\scriptscriptstyle-.15 }$\\
  M3-33    &             &                                                                  & 0.917  $\pm.084$&  2.93$\stackrel{\scriptscriptstyle+.14}{\scriptscriptstyle-.17 }$&                &                                                                  &   0.980$\pm.029$&  3.49$\stackrel{\scriptscriptstyle+.04}{\scriptscriptstyle-.04 }$\\
  M3-7     &             &                                                                  & 0.556  $\pm.015$&  3.79$\stackrel{\scriptscriptstyle+.08}{\scriptscriptstyle-.07 }$& 0.990 $\pm.196$&  3.30$\stackrel{\scriptscriptstyle+.24}{\scriptscriptstyle-.45 }$&                 &                                                                  \\
  Me1-1    &             &                                                                  & 0.546  $\pm.021$&  3.83$\stackrel{\scriptscriptstyle+.10}{\scriptscriptstyle-.09 }$& 0.741 $\pm.066$&  3.66$\stackrel{\scriptscriptstyle+.09}{\scriptscriptstyle-.10 }$&   0.735$\pm.016$&  3.84$\stackrel{\scriptscriptstyle+.03}{\scriptscriptstyle-.03 }$\\
  Me2-1    &             &                                                                  &                 &                                                                  &                &                                                                  &   1.111$\pm.030$&  3.26$\stackrel{\scriptscriptstyle+.06}{\scriptscriptstyle-.06 }$\\
  Me2-2    &.409$\pm.008$&  4.07$\stackrel{\scriptscriptstyle+.06}{\scriptscriptstyle-.06 }$&                 &                                                                  &                &                                                                  &                 &                                                                  \\   
  MyCn18   &              &                                                                 & 0.552  $\pm.015$&  3.80$\stackrel{\scriptscriptstyle+.07}{\scriptscriptstyle-.07 }$& 0.485 $\pm.054$&  4.08$\stackrel{\scriptscriptstyle+.11}{\scriptscriptstyle-.12 }$&   1.119$\pm.143$&  3.25$\stackrel{\scriptscriptstyle+.23}{\scriptscriptstyle-.47 }$\\ 
  Mz3$^s$  &              &                                                                 & 0.571  $\pm.013$&  3.72$\stackrel{\scriptscriptstyle+.05}{\scriptscriptstyle-.05 }$& 0.333 $\pm.063$&  4.49$\stackrel{\scriptscriptstyle+.26}{\scriptscriptstyle-.24 }$&                 &                                                                   \\           
  NGC3242  &              &                                                                 & 0.725  $\pm.042$&  3.30$\stackrel{\scriptscriptstyle+.09}{\scriptscriptstyle-.09 }$& 1.082 $\pm.069$&  3.15$\stackrel{\scriptscriptstyle+.10}{\scriptscriptstyle-.14 }$&   1.079$\pm.022$&  3.32$\stackrel{\scriptscriptstyle+.04}{\scriptscriptstyle-.04 }$\\
  NGC40    &.750$\pm.032$&  3.10$\stackrel{\scriptscriptstyle+.05}{\scriptscriptstyle-.05 }$& 0.725  $\pm.058$&  3.29$\stackrel{\scriptscriptstyle+.12}{\scriptscriptstyle-.12 }$& 1.186 $\pm.042$&  2.95$\stackrel{\scriptscriptstyle+.08}{\scriptscriptstyle-.10 }$&                 &                                                                  \\
  NGC5307  &             &                                                                  & 0.625  $\pm.055$&  3.55$\stackrel{\scriptscriptstyle+.17}{\scriptscriptstyle-.16 }$& 1.464 $\pm.219$&  $<$1.15                                                         &   1.059$\pm.019$&  3.36$\stackrel{\scriptscriptstyle+.03}{\scriptscriptstyle-.04 }$\\
  NGC5873  &.511$\pm.023$&  3.61$\stackrel{\scriptscriptstyle+.08}{\scriptscriptstyle-.07 }$& 0.571  $\pm.020$&  3.72$\stackrel{\scriptscriptstyle+.09}{\scriptscriptstyle-.08 }$& 0.813 $\pm.093$&  3.55$\stackrel{\scriptscriptstyle+.12}{\scriptscriptstyle-.15 }$&   0.943$\pm.018$&  3.55$\stackrel{\scriptscriptstyle+.02}{\scriptscriptstyle-.02 }$\\
  NGC5882  &.485$\pm.026$&  3.70$\stackrel{\scriptscriptstyle+.10}{\scriptscriptstyle-.09 }$&                 &                                                                  &                &                                                                  &                 &                                                                  \\
  NGC5979  &.690$\pm.038$&  3.21$\stackrel{\scriptscriptstyle+.07}{\scriptscriptstyle-.07 }$& 0.775  $\pm.090$&  3.19$\stackrel{\scriptscriptstyle+.17}{\scriptscriptstyle-.19 }$& 0.917 $\pm.135$&  3.41$\stackrel{\scriptscriptstyle+.16}{\scriptscriptstyle-.23 }$&   1.107$\pm.018$&  3.27$\stackrel{\scriptscriptstyle+.03}{\scriptscriptstyle-.04 }$\\
  NGC6058  &.722$\pm.130$&  3.15$\stackrel{\scriptscriptstyle+.25}{\scriptscriptstyle-.21 }$&                 &                                                                  &                &                                                                  &   1.229$\pm.208$&  2.96$\stackrel{\scriptscriptstyle+.41}{\scriptscriptstyle-1.7}$\\
  NGC6072  &             &                                                                  & 1.103  $\pm.030$&  2.59$\stackrel{\scriptscriptstyle+.06}{\scriptscriptstyle-.06 }$& 1.508 $\pm.284$&  $<$1.15                                                                   &      $\pm.043$&  $<$1.26                                                       \\
  NGC6153  &.474$\pm.002$&  3.74$\stackrel{\scriptscriptstyle+.01}{\scriptscriptstyle-.01 }$& 0.617  $\pm.023$&  3.56$\stackrel{\scriptscriptstyle+.07}{\scriptscriptstyle-.07 }$& 0.794 $\pm.101$&  3.58$\stackrel{\scriptscriptstyle+.13}{\scriptscriptstyle-.17 }$&   0.962$\pm.028$&  3.52$\stackrel{\scriptscriptstyle+.05}{\scriptscriptstyle-.05 }$\\
  NGC6210  &.509$\pm.021$&  3.62$\stackrel{\scriptscriptstyle+.07}{\scriptscriptstyle-.06 }$& 0.602  $\pm.015$&  3.61$\stackrel{\scriptscriptstyle+.04}{\scriptscriptstyle-.04 }$& 0.855 $\pm.015$&  3.50$\stackrel{\scriptscriptstyle+.02}{\scriptscriptstyle-.02 }$&   0.763$\pm.274$&  3.80$\stackrel{\scriptscriptstyle+.28}{\scriptscriptstyle-.73 }$\\
  NGC6302  &.460$\pm.024$&  3.80$\stackrel{\scriptscriptstyle+.11}{\scriptscriptstyle-.09 }$& 0.500  $\pm.020$&  4.11$\stackrel{\scriptscriptstyle+.19}{\scriptscriptstyle-.15 }$& 0.369 $\pm.031$&  4.36$\stackrel{\scriptscriptstyle+.10}{\scriptscriptstyle-.10 }$&   0.559$\pm.006$&  4.09$\stackrel{\scriptscriptstyle+.01}{\scriptscriptstyle-.01 }$\\
  NGC6439  &.527$\pm.013$&  3.56$\stackrel{\scriptscriptstyle+.04}{\scriptscriptstyle-.04 }$& 0.568  $\pm.026$&  3.74$\stackrel{\scriptscriptstyle+.10}{\scriptscriptstyle-.09 }$& 0.685 $\pm.061$&  3.74$\stackrel{\scriptscriptstyle+.09}{\scriptscriptstyle-.10 }$&   0.763$\pm.017$&  3.80$\stackrel{\scriptscriptstyle+.02}{\scriptscriptstyle-.02 }$\\
  NGC6537  &             &                                                                  & 0.505  $\pm.008$&  4.07$\stackrel{\scriptscriptstyle+.06}{\scriptscriptstyle-.06 }$& 0.407 $\pm.038$&  4.26$\stackrel{\scriptscriptstyle+.10}{\scriptscriptstyle-.10 }$&   0.645$\pm.046$&  3.97$\stackrel{\scriptscriptstyle+.06}{\scriptscriptstyle-.07 }$\\
  NGC6543  &.491$\pm.010$&  3.68$\stackrel{\scriptscriptstyle+.03}{\scriptscriptstyle-.03 }$&                 &                                                                  &                &                                                                  &                 &                                                                  \\
  NGC6565  &             &                                                                  & 0.741  $\pm.022$&  3.26$\stackrel{\scriptscriptstyle+.05}{\scriptscriptstyle-.05 }$& 1.083 $\pm.117$&  3.15$\stackrel{\scriptscriptstyle+.17}{\scriptscriptstyle-.26 }$&   1.266$\pm.064$&  2.84$\stackrel{\scriptscriptstyle+.20}{\scriptscriptstyle-.37 }$\\
 NGC6567$^s$&.434$\pm.008$&  3.92$\stackrel{\scriptscriptstyle+.05}{\scriptscriptstyle-.04 }$& 0.559  $\pm.019$&  3.78$\stackrel{\scriptscriptstyle+.09}{\scriptscriptstyle-.08 }$& 0.505 $\pm.048$&  4.04$\stackrel{\scriptscriptstyle+.09}{\scriptscriptstyle-.10 }$&   0.690$\pm.015$&  3.90$\stackrel{\scriptscriptstyle+.02}{\scriptscriptstyle-.02 }$\\
  NGC6572  &.421$\pm.024$&  3.99$\stackrel{\scriptscriptstyle+.18}{\scriptscriptstyle-.13 }$& 0.472  $\pm.016$&  4.42$\stackrel{\scriptscriptstyle+2.5}{\scriptscriptstyle-.21 }$& 0.439 $\pm.015$&  4.18$\stackrel{\scriptscriptstyle+.03}{\scriptscriptstyle-.04 }$&   0.500$\pm.020$&  4.18$\stackrel{\scriptscriptstyle+.03}{\scriptscriptstyle-.03 }$\\
  NGC6578  &             &                                                                  & 0.625  $\pm.020$&  3.54$\stackrel{\scriptscriptstyle+.06}{\scriptscriptstyle-.06 }$& 0.971 $\pm.132$&  3.33$\stackrel{\scriptscriptstyle+.17}{\scriptscriptstyle-.26 }$&   0.901$\pm.024$&  3.61$\stackrel{\scriptscriptstyle+.06}{\scriptscriptstyle-.06 }$\\
  NGC6620  &.605$\pm.012$&  3.37$\stackrel{\scriptscriptstyle+.03}{\scriptscriptstyle-.02 }$& 0.671  $\pm.032$&  3.42$\stackrel{\scriptscriptstyle+.08}{\scriptscriptstyle-.08 }$& 0.901 $\pm.065$&  3.43$\stackrel{\scriptscriptstyle+.09}{\scriptscriptstyle-.11 }$&   1.030$\pm.031$&  3.41$\stackrel{\scriptscriptstyle+.05}{\scriptscriptstyle-.06 }$\\
  NGC6720  &.995$\pm.044$&  2.72$\stackrel{\scriptscriptstyle+.07}{\scriptscriptstyle-.07 }$& 1.036  $\pm.046$&  2.72$\stackrel{\scriptscriptstyle+.08}{\scriptscriptstyle-.09 }$& 1.287 $\pm.045$&  2.69$\stackrel{\scriptscriptstyle+.13}{\scriptscriptstyle-.17 }$&   1.248$\pm.061$&  2.90$\stackrel{\scriptscriptstyle+.18}{\scriptscriptstyle-.30 }$\\
  NGC6741  &.507$\pm.022$&  3.62$\stackrel{\scriptscriptstyle+.07}{\scriptscriptstyle-.07 }$& 0.571  $\pm.010$&  3.72$\stackrel{\scriptscriptstyle+.04}{\scriptscriptstyle-.04 }$& 0.725 $\pm.032$&  3.68$\stackrel{\scriptscriptstyle+.04}{\scriptscriptstyle-.05 }$&   0.735$\pm.043$&  3.84$\stackrel{\scriptscriptstyle+.06}{\scriptscriptstyle-.07 }$\\
  NGC6778  &.812$\pm.021$&  3.00$\stackrel{\scriptscriptstyle+.03}{\scriptscriptstyle-.03 }$& 0.893  $\pm.016$&  2.97$\stackrel{\scriptscriptstyle+.03}{\scriptscriptstyle-.03 }$& 0.893 $\pm.016$&  3.44$\stackrel{\scriptscriptstyle+.02}{\scriptscriptstyle-.02 }$&                 &                                                                  \\
  NGC6781  &1.17$\pm0.04$&  2.43$\stackrel{\scriptscriptstyle+.07}{\scriptscriptstyle-.07 }$& 1.214  $\pm.028$&  2.35$\stackrel{\scriptscriptstyle+.07}{\scriptscriptstyle-.08 }$& 1.126 $\pm.169$&  3.07$\stackrel{\scriptscriptstyle+.24}{\scriptscriptstyle-.52 }$&   2.387$\pm.655$&  $<$1.26                                                         \\
  NGC6790  &.399$\pm.023$&  4.15$\stackrel{\scriptscriptstyle+.25}{\scriptscriptstyle-.17 }$& 0.452  $\pm.000$&  5.04$\stackrel{\scriptscriptstyle+.00}{\scriptscriptstyle-.00 }$& 0.402 $\pm.000$&  4.27$\stackrel{\scriptscriptstyle+.00}{\scriptscriptstyle-.00 }$&   0.242$\pm.001$&  4.78$\stackrel{\scriptscriptstyle+.00}{\scriptscriptstyle-.00 }$\\
  NGC6803  &.477$\pm.005$&  3.73$\stackrel{\scriptscriptstyle+.02}{\scriptscriptstyle-.02 }$&                 &                                                                  &                &                                                                  &                 &                                                                  \\
  NGC6807  &.434$\pm.010$&  3.92$\stackrel{\scriptscriptstyle+.06}{\scriptscriptstyle-.05 }$&                 &                                                                  &                &                                                                  &                 &                                                                  \\
  NGC6818  &.682$\pm.013$&  3.22$\stackrel{\scriptscriptstyle+.02}{\scriptscriptstyle-.02 }$& 0.741  $\pm.016$&  3.27$\stackrel{\scriptscriptstyle+.04}{\scriptscriptstyle-.04 }$& 1.029 $\pm.043$&  3.24$\stackrel{\scriptscriptstyle+.06}{\scriptscriptstyle-.07 }$&   1.070$\pm.014$&  3.34$\stackrel{\scriptscriptstyle+.02}{\scriptscriptstyle-.03 }$\\
  NGC6826  &.657$\pm.031$&  3.27$\stackrel{\scriptscriptstyle+.06}{\scriptscriptstyle-.06 }$& 0.730  $\pm.027$&  3.28$\stackrel{\scriptscriptstyle+.06}{\scriptscriptstyle-.06 }$& 1.098 $\pm.052$&  3.12$\stackrel{\scriptscriptstyle+.08}{\scriptscriptstyle-.10 }$&   1.073$\pm.092$&  3.33$\stackrel{\scriptscriptstyle+.15}{\scriptscriptstyle-.21 }$\\
  \noalign{\smallskip}                                                                                                                                                                                                                                                                
  \hline                                                                                                                                                                                                                                                                              
 \end{tabular}                                                                                                                                                                                                                                                                       
\end{table*}                                                                                                                                                                                                                                                                         
                                                                                                                                                                                                                                                                                     
\setcounter{table}{1}                                                                                                                                                                                                                                                                
\begin{table*}                                                                                                                                                                                                                                                                       
\caption{ {  \it --continued} }                                                                                                                                                                                                                                                      
\centering                                                                                                                                                                                                                                                                           
\begin{tabular}{l c c c c c c c c c c}                                                                                                                                                                                                                                               
\hline                                                                                                                                                                                                                                                                               
\hline                                                                                                                                                                                                                                                                               
\noalign{\smallskip}                                                                                                                                                                                                                                                                 
       & \multicolumn{2}{c}{[OII]}  & \multicolumn{2}{c}{[SII]} & \multicolumn{2}{c}{[ClIII]} & \multicolumn{2}{c}{[ArIV]}\\                                                                                                                                                         
%Source & $\lambda3729/$ & log$\mathrm {Ne}$ & $\lambda6731/$ & log$\mathrm {Ne}$ &$\lambda5537/$ & log$\mathrm {Ne}$ & $\lambda4740/$ & Log$\rm {N                                                                                                                                  
Source & $\lambda3729/$ & ${\rm Log} N_{\rm e}$ & $\lambda6716/$ & ${\rm Log} N_{\rm e}$ &$\lambda5517/$ & ${\rm Log} N_{\rm e}$ & $\lambda4711/$ & ${\rm Log} N_{\rm e}$\\                                                                                                          
&$\lambda3726$ & (cm$^{-3}$) &$\lambda6731$ & (cm$^{-3}$) &  $\lambda5537$ & (cm$^{-3}$) & $\lambda4740$ & (cm$^{-3}$)&\\                                                                                                                                                            
\noalign{\smallskip}                                                                                                                                                                                                                                                                 
\hline                                                                                                                                                                                                                                                                               
                                                                                                                                                                                                                                                                                     
  NGC6833  &.381$\pm.014$&  4.34$\stackrel{\scriptscriptstyle+.22}{\scriptscriptstyle-.15 }$&                  &                                                                  &                 &                                                                  &                       &                                                                  \\ 
  NGC6853  &1.19$\pm0.02$&  2.40$\stackrel{\scriptscriptstyle+.03}{\scriptscriptstyle-.03 }$&                  &                                                                  &                 &                                                                  &                       &                                                                  \\
  NGC6879  &.492$\pm.007$&  3.67$\stackrel{\scriptscriptstyle+.03}{\scriptscriptstyle-.03 }$&                  &                                                                  &                 &                                                                  &                       &                                                                  \\
  NGC6884  &.465$\pm.037$&  3.77$\stackrel{\scriptscriptstyle+.17}{\scriptscriptstyle-.13 }$&  0.529  $\pm.020$&  3.91$\stackrel{\scriptscriptstyle+.11}{\scriptscriptstyle-.10 }$&  0.699$\pm.024$&  3.72$\stackrel{\scriptscriptstyle+.04}{\scriptscriptstyle-.04 }$&  0.606$\pm.206$&  4.02$\stackrel{\scriptscriptstyle+.24}{\scriptscriptstyle-.43 }$\\
  NGC6886  &.501$\pm.007$&  3.64$\stackrel{\scriptscriptstyle+.02}{\scriptscriptstyle-.02 }$&                  &                                                                  &                &                                                                  &                &                                                                  \\
  NGC7008  &.770$\pm.038$&  3.07$\stackrel{\scriptscriptstyle+.06}{\scriptscriptstyle-.06 }$&                  &                                                                  &                &                                                                  &                &                                                                  \\
  NGC7009  &.498$\pm.019$&  3.65$\stackrel{\scriptscriptstyle+.07}{\scriptscriptstyle-.06 }$&  0.599  $\pm.022$&  3.62$\stackrel{\scriptscriptstyle+.07}{\scriptscriptstyle-.07 }$&  0.787$\pm.031$&  3.59$\stackrel{\scriptscriptstyle+.04}{\scriptscriptstyle-.05 }$&  0.885$\pm.016$&  3.63$\stackrel{\scriptscriptstyle+.02}{\scriptscriptstyle-.03 }$\\
  NGC7026  &.549$\pm.010$&  3.50$\stackrel{\scriptscriptstyle+.03}{\scriptscriptstyle-.02 }$&                  &                                                                  &                &      &                                                           &                &                                                                  \\
  NGC7662  &.605$\pm.031$&  3.37$\stackrel{\scriptscriptstyle+.07}{\scriptscriptstyle-.06 }$&  0.654  $\pm.026$&  3.46$\stackrel{\scriptscriptstyle+.07}{\scriptscriptstyle-.07 }$&  1.010$\pm.034$&  3.27$\stackrel{\scriptscriptstyle+.05}{\scriptscriptstyle-.05 }$&  1.065$\pm.025$&  3.35$\stackrel{\scriptscriptstyle+.04}{\scriptscriptstyle+.05 }$\\
  PC12     &             &                                                                  &  0.568  $\pm.016$&  3.73$\stackrel{\scriptscriptstyle+.06}{\scriptscriptstyle-.06 }$&  0.758$\pm.092$&  3.63$\stackrel{\scriptscriptstyle+.12}{\scriptscriptstyle-.15 }$&                &                                                                 \\
  PC14     &             &                                                                  &  0.625  $\pm.020$&  3.54$\stackrel{\scriptscriptstyle+.06}{\scriptscriptstyle-.06 }$&  0.813$\pm.079$&  3.55$\stackrel{\scriptscriptstyle+.11}{\scriptscriptstyle-.13 }$&  0.926$\pm.026$&  3.58$\stackrel{\scriptscriptstyle+.03}{\scriptscriptstyle-.04 }$\\
  PRMG1    &             &                                                                  &                  &                                                                  &                &                                                                  &  1.220$\pm.073$&  3.00$\stackrel{\scriptscriptstyle+.18}{\scriptscriptstyle+.31 }$\\
  Sn1      &.632$\pm.037$&  3.32$\stackrel{\scriptscriptstyle+.08}{\scriptscriptstyle-.07 }$&  0.490  $\pm.065$&  4.21$\stackrel{\scriptscriptstyle+1.3}{\scriptscriptstyle-.45 }$&  0.990$\pm.255$&  3.30$\stackrel{\scriptscriptstyle+.29}{\scriptscriptstyle-.78 }$&  1.066$\pm.047$&  3.35$\stackrel{\scriptscriptstyle+.08}{\scriptscriptstyle+.10 }$\\
  Vy1-1    &.613$\pm.021$&  3.36$\stackrel{\scriptscriptstyle+.05}{\scriptscriptstyle-.04 }$&                  &                                                                  &                &                                                                  &                &                                                                  \\
  Vy2-1    &             &                                                                  &  0.621  $\pm.027$&  3.55$\stackrel{\scriptscriptstyle+.08}{\scriptscriptstyle-.08 }$&  0.714$\pm.061$&  3.69$\stackrel{\scriptscriptstyle+.08}{\scriptscriptstyle-.09 }$&  0.917$\pm.042$&  3.59$\stackrel{\scriptscriptstyle+.06}{\scriptscriptstyle-.07 }$\\
                                                                                                                                                                                                                                                                                                       
\noalign{\smallskip}                                                                                                                                                                                                                                                                                                                                                            
\hline 
\end{tabular}
\begin{list}{}{}
\item $^a$ A superscript ``$^s$'' indicates that the measured line ratios were based on scanned spectra.
\end{list}                                                                                                                                                                                                                                                                                                                                                                           
\end{table*}

%% file: Table03.tex
\begin{table}
 \caption{\label{atomic_ref}References for default atomic parameters}
 \begin{center}
 \begin{tabular}{ccccc}
 \hline
 \hline
 \noalign{\smallskip}
 Ion & \multicolumn{2}{c}{IP\,(eV)} & \multicolumn{2}{c}{Reference}\\
\cline{2-3}
\cline{4-5}
\noalign{\smallskip}
X$^i$& X$^{i-1}$ & X$^i$  & Trans. Prob. & Coll. Str.  \\
 \noalign{\smallskip}
 \hline
 \noalign{\smallskip}
  O$^+$& 13.62 & 35.12 & [1] & [2] \\
  S$^+$& 10.36 & 23.34 & [3],[4] & [5] \\
 Cl$^{2+}$& 23.8 & 39.9  & [3]  & [6]  \\
 Ar$^{3+}$& 40.74 & 59.81 & [3]  & [7]  \\
 \noalign{\smallskip}
 \hline
 \end{tabular}
 \end{center}
References: [1] Zeippen (\cite{zeippen1982}); [2] Pradhan
(\cite{pradhan}); [3] Mendoza \& Zeippen (\cite{mendoza1982}); [4] Keenan et
al. (\cite{keenan1993}); [5] Keenan et al. (\cite{keenan1996}); [6] Butler \&
Zeippen (\cite{butler1989}); [7] Ramsbottom, Bell, \& Keenan (\cite{Ramsbottom})\end{table}

%% file: Table04.tex
\begin{table}
\caption{\label{fig7_ref} References for O$^+$ atomic data used to
generate the loci plotted in Fig.\,7. }
 \begin{center}
 \begin{tabular}{ccc}
 \hline
 \hline
 \noalign{\smallskip}
 Line &  \multicolumn{2}{c}{Reference}\\
\cline{2-3}
\noalign{\smallskip}
 &  Trans. Prob. & Coll. Str.  \\
 \noalign{\smallskip}
 \hline
 \noalign{\smallskip}
 dotted-dotted-dashed & [2] & [4]\\
 solid & [1] & [3]\\
 short-dashed   & [2] & [3]\\
 long-dashed  & [5] & [3]\\
\noalign{\smallskip}
 \hline
 \end{tabular}
 \end{center}
References: [1] Zeippen (\cite{zeippen1982});  [2] Zeippen
(\cite{zeippen1987a});  [3] Pradhan (\cite{pradhan});  [4] McLaughlin \&
Bell (\cite{mclaughlin}) ;  [5] Wiese et al. (\cite{wiese})
\end{table}

%% file: Table05.tex
  \begin{table}
\caption{[\ion{O}{ii}] $\lambda3729/\lambda3726$ intensity ratios 
of low-$N_{\rm e}$ plasmas from literature }
\label{oiirat}
\centering
\begin{tabular}{lccc}
\hline
\hline
\noalign{\smallskip}
Source & obs.pos & $I(\lambda3729)/I(\lambda3726)$ & Ref. \\
\noalign{\smallskip}
\hline
\noalign{\smallskip}
%$0.74\pm0.67$ & $1.79\stackrel{\scriptscriptstyle+1.04}{\scriptscriptstyle-0.69}$
% IC418& 180" & $0.74\pm0.67$ & $1.79\stackrel{2.83}{1.30}$&MBC\\
% IC418&31.4" & $3.87\pm0.31$ & $0.34\stackrel{0.37}{0.31}$&MBC\\
 IC418      & 180$^{\prime\prime} $  & $1.79\stackrel{\scriptscriptstyle2.83}{\scriptscriptstyle1.30}$ &[1]\\
% IC418      & 31.4$\prime\prime$  & $0.34\stackrel{\scriptscriptstyle0.37}{\scriptscriptstyle0.31}$ &[1]\\
 PX Pup     & 14$^{\prime\prime}$    & $1.43\stackrel{\scriptscriptstyle1.54}{\scriptscriptstyle1.33}$ &[1]\\
%NGC 6072   & Diff. Em. & $1.39\stackrel{\scriptscriptstyle1.57}{\scriptscriptstyle1.21}$ &[2]\\
 He 2-146   &        & $1.43\stackrel{\scriptscriptstyle1.45}{\scriptscriptstyle1.41}$ &[2]\\
 NGC 7293   &        & $1.41\stackrel{\scriptscriptstyle1.43}{\scriptscriptstyle1.39}$ &[2]\\
 IC 5148-50 &        & $1.45\stackrel{\scriptscriptstyle1.49}{\scriptscriptstyle1.41}$ &[2]\\
 NGC6543    &halo    & $1.44\stackrel{\scriptscriptstyle1.56}{\scriptscriptstyle1.32}$ &[3]\\
 NGC 2371-2 & 9$^{\prime\prime}$     & $1.47\stackrel{\scriptscriptstyle1.73}{\scriptscriptstyle1.21}$& [4]\\  
 %NGC6826    &halo    & $1.07\stackrel{\scriptscriptstyle1.57}{\scriptscriptstyle0.57}$ &[3]\\
% NGC 650    & 20$\prime\prime$    &  1.40                                                           &[4]\\            
% M 2-51     &        &  1.63                                                           &[4]\\
% NGC 2452   & SE     &  1.66                                                           &[5]\\  
% NGC 7026   &        &  1.62                                                           &[5]\\
% NGC 7026   &        &  1.68                                                           &[6]\\ 
\hline
\end{tabular}
 \begin{list}{}{}
 \item[References:] 
 [1] Monk, Barlow \& Clegg (\cite{MBC1990});
 [2] Kingsburgh \& English (\cite{kingsburgh1992});
 [3] Middlemass, Clegg \& Walsh (\cite{MCW1989});
 [4] Kingsburgh \& Barlow (\cite{KB92})
% [4]  Kaler,Shaw,\&Kwitter(\cite{KSK1990});
% [5]  Aller, \& Czyzak (\cite{AC1979});
% [6]  Czyzak, \& Aller(\cite{CA1970})
\end{list}
\end{table}

%% file: Table06.tex
  \begin{table}
\caption{Extended diffuse emission regions around PNe that give a [\ion{S}{ii}] 
$\lambda6716$/$\lambda6731$ ratio larger than 1.4 (from Kingsburgh \& English 
1992).}
\label{siirat}
\centering
\begin{tabular}{lc}
\hline
\hline
\noalign{\smallskip}
PN & $I(\lambda6716)/I(\lambda6731)^a$\\
\noalign{\smallskip}
\hline
\noalign{\smallskip}
 NGC 3195   &  1.98::                                                         \\
 Fg 1       & $1.45\stackrel{\scriptscriptstyle1.56}{\scriptscriptstyle1.34}$ \\
 Pb 8       & $1.40\stackrel{\scriptscriptstyle1.48}{\scriptscriptstyle1.32}$ \\
 He 2-104   & $1.71\stackrel{\scriptscriptstyle1.92}{\scriptscriptstyle1.50}$ \\
 He 2-141   & $1.62\stackrel{\scriptscriptstyle1.84}{\scriptscriptstyle1.40}$ \\
 He 2-146   & $1.52\stackrel{\scriptscriptstyle1.67}{\scriptscriptstyle1.37}$ \\            
 NGC 6072   & $1.61\stackrel{\scriptscriptstyle1.80}{\scriptscriptstyle1.42}$ \\  
 Mz 2       & $1.44\stackrel{\scriptscriptstyle1.55}{\scriptscriptstyle1.33}$ \\
 IC 4634    &  2.14::                                                         \\ 
 IC 4637    &  1.75::                                                         \\
\noalign{\smallskip}
\hline
\end{tabular}
 \begin{list}{}{}
\item[$^a$] A "::" indicates an uncertainty larger than 20\%.
\end{list} 
\end{table}

%% file: Table07.tex
\begin{table}
\caption{\label {kstest} Results of Kolmogorov-Smirnov Two-sample Test on
densities deduced from the four optical $N_{\rm e}$-diagnostic ratios. 
X and Y denote the two density samples for the test, $D$ is the
Kolmogorv-Smirnov D statistic of the samples, $P_0(>|D|)$ gives the probability 
of exceeding $|D|$ under the Null Hypothesis and is equal to 1 minus confidence 
factor, $n$ is the size of the samples and $D(\alpha _{0.10})$ is the 
critical $D$ value for a confidence factor of 0.10.}
\centering
\begin{tabular}{ccrccc}
\hline
\hline
\noalign{\smallskip}
 X & Y & $D$\hspace{4mm} & $P_0(>|D|)$ & $n$ & $D(\alpha _{0.10})$ \\
\hline
\noalign{\smallskip}
[\ion{O}{II}]& [\ion{S}{II}] & 0.162 & 0.715 & 37 &0.284\\

[\ion{O}{II}]& [\ion{Cl}{III}] &$-$0.139 & 0.878 &36 &0.288\\

[\ion{O}{II}]& [\ion{Ar}{IV}]  &0.245 & 0.106 &49 &0.246\\

[\ion{S}{II}]& [\ion{Ar}{IV}]  &$-$0.135 & 0.509 &62 & 0.200\\

[\ion{S}{II}]& [\ion{Cl}{III}] &$-$0.113 & 0.824 &74 & 0.219\\

[\ion{Ar}{IV}]& [\ion{Cl}{III}] &0.152 & 0.498 &59&0.225 \\
\noalign{\smallskip}
 \hline
 \end{tabular}
\end{table}